\documentclass[10pt]{article}

\usepackage[margin=1in]{geometry}
\usepackage{amsmath,amssymb,amsfonts}
\usepackage{mathtools}
\usepackage{graphicx}
\usepackage{xcolor}
\usepackage{booktabs}
\usepackage{multirow}
\usepackage{float}
\usepackage{enumitem}
\usepackage{url}
\usepackage{tikz}
\usepackage{tikz-cd}
\usepackage{algorithm}
\usepackage{algpseudocode}

\usepackage{bbm} 
\usepackage{physics}         
\usepackage{subfig}          
\usepackage{mdframed}        

\newenvironment{boxwithhead}[1][]{%
  \begin{mdframed}[linewidth=0.5pt]\noindent\textbf{#1}\par\medskip
}{%
  \end{mdframed}
}
\usepackage{fvextra}
 \DefineVerbatimEnvironment{codeblock}{Verbatim}{
   frame=single,
   framesep=3mm,
   fontsize=\small,
   breaklines=true,
   commandchars=\\\{\}
}

\newcommand{\bmsubsubsection}[1]{\subsubsection{#1}}
\newcommand{\E}{\mathbb{E}}
\newcommand{\dist}{\operatorname{dist}}
\newcommand{\Var}{\operatorname{Var}}
\usepackage[numbers]{natbib}

\usepackage[hidelinks]{hyperref}
\usepackage[capitalize,nameinlink]{cleveref}

\usepackage{amsthm}

\newtheorem{remark}{Remark}[section]
\theoremstyle{remark}

\usepackage{authblk}

\title{\textbf{Feasible Dose–Response Curves for Continuous Treatments Under Positivity Violations}}

\author[1]{Han Bao\thanks{Corresponding author. Email: \texttt{han.bao@stat.uni-muenchen.de}}}
\author[1,2,3,4]{Michael Schomaker}

\affil[1]{Department of Statistics, Ludwig-Maximilians-Universit\"at M\"unchen, Munich, Germany}
\affil[2]{Centre for Integrated Data and Epidemiological Research, Cape Town, South Africa}
\affil[3]{Institute of Public Health, Medical Decision Making and Health Technology Assessment,\\
UMIT -- University for Health Sciences, Medical Informatics and Technology, Hall in Tirol, Austria}
\affil[4]{Munich Center for Machine Learning (MCML), Ludwig-Maximilians-Universit\"at M\"unchen, Munich, Germany}

\date{} 

\newcommand{\keywords}[1]{\par\smallskip\noindent\textbf{Keywords: }#1\par}

\begin{document}
\maketitle

\begin{abstract}


Positivity violations can complicate estimation and interpretation of causal dose–response curves (CDRCs) for continuous interventions. Recently proposed weighting-based methods are designed to handle limited overlap, but the resulting weighted targets can be difficult to interpret scientifically. Modified treatment policies (MTPs) are often less sensitive to support limitations if designed properly, yet they typically target policy-defined effects that may not align with the original dose–response question. We develop an approach that addresses limited overlap while remaining as close as possible to the scientific target of the CDRC.

Our work is motivated by the CHAPAS-3 trial of HIV-positive children in Zambia and Uganda, where clinically relevant efavirenz concentration levels are not uniformly supported across covariate strata. We introduce a new diagnostic—the non-overlap ratio—which, as a function of the target intervention level, quantifies the proportion of the population for whom that level is not supported given observed covariates. We also define an individualized most feasible intervention: for each child and target concentration, we retain the target when it is supported, and otherwise map it to the nearest supported concentration. The resulting feasible dose–response curve answers: if we try to set everyone to a given concentration, but that concentration is not realistically attainable for some individuals, what outcome would be expected after shifting those individuals to their nearest attainable concentration?

We propose a plug-in g-computation estimator that combines outcome regression with flexible conditional density estimation to learn supported regions and evaluate the feasible estimand. In simulations, the method reduces bias under positivity violations and recovers the standard CDRC when support is adequate. We apply the approach to CHAPAS-3, demonstrating that feasible dose–response estimation yields a stable and interpretable concentration–response summary under realistic support constraints.
\end{abstract}


\section{Introduction}\label{sec:int}
Causal inference methods have been extensively studied for binary treatments, where individuals are either exposed to a treatment or assigned to a control condition \cite{imbens2015causal, hernan2020causal}. In this framework, the positivity assumption—requiring that each individual has a non-zero probability of receiving any treatment level under consideration within confounder strata—is crucial for the identifiability and estimation of causal effects.

Violations of the positivity assumption can be categorized as structural or practical \cite{zivich2022positivity, petersen2012diagnosing, westreich2010invited}. Structural violations occur when an intervention is inherently impossible, whereas practical violations arise when certain treatment levels are theoretically possible but are not observed in the data. Both types lead to non-overlap in the confounder space, where certain confounder patterns have a zero probability—either theoretical or estimated—of receiving specific treatment levels. These violations hinder the estimation of causal effects, making it challenging to draw reliable inferences.

In pharmacoepidemiological and public health studies, researchers often extend their focus to continuous interventions, where the treatment variable can take on a range of values. In this context, the estimation of the causal dose–response curve (CDRC)—which describes the relationship between the continuous treatment and the outcome—has gained increasing attention in the causal inference literature. The CDRC provides a nuanced understanding of the causal effects of continuous interventions. However, estimating the CDRC introduces new challenges, particularly because the identification of the CDRC necessitates a more restrictive positivity assumption.

In this paper, ``dose'' is used generically to denote a continuous intervention level. Accordingly, references to the dose--response curve also encompass concentration--response and other continuous-exposure response relations.

For CDRC estimation, positivity can be more difficult to satisfy in practice because the treatment is continuous and support must be assessed over a much richer intervention space. In finite samples, some treatment regions may be sparsely represented within certain confounder strata, even when positivity is plausible at the population level. This limited overlap can reduce precision and, depending on the estimator and degree of extrapolation, may also introduce bias.

While methods adapted from binary treatments can be applied, they have limitations. Trimming, for example, excludes regions of the confounder space with poor overlap \cite{crump2009dealing}, removing individuals with extreme probabilities of treatment given their confounders. However, this approach may result in the loss of valuable data. Moreover, it changes the target estimand: the resulting effect is interpreted conditional on the retained subset of the population, rather than the original full study population.

To tackle these challenges, researchers have developed specialized methods, including projection functions, which aim to mitigate the violation of the positivity assumption \cite{petersen2012diagnosing}. One such approach is the weighted CDRC, which reweights and redefines the estimand of interest based on functions of the conditional support for the respective interventions \cite{Schomaker:2024}. Additionally, machine learning techniques have proven effective in enhancing estimation accuracy by minimizing the risk of bias caused by model misspecification \cite{moccia2024machine}. 

A related approach is to consider modified treatment policies (MTPs) \cite{haneuse2013estimation, diaz2023nonparametric}, which define scientifically motivated interventions by systematically transforming each individual’s observed (natural) treatment (e.g., shifting, or truncating doses). Because MTPs map observed treatment values to nearby, well-supported values, they can also improve overlap and mitigate practical positivity concerns as a secondary benefit. When designed carefully, MTPs can substantially reduce reliance on the positivity assumption. However, they typically target a different causal estimand than the causal dose–response curve (CDRC), which may or may not align with the scientific objectives of a particular study—including the empirical example discussed in Section 5. MTPs are most useful when the modified intervention itself is of substantive scientific interest. Haneuse and Rotnitzky provide an example of a very well-motivated MTP \cite{haneuse2013estimation}.

While double robustness and semiparametric efficiency are often desirable, they are not generally available for estimation of the full CDRC under continuous interventions. Under mild smoothness conditions, the full CDRC functional is not pathwise differentiable \cite{diaz2013targeted}, which precludes regular root-\(n\)-consistent estimation and corresponding first-order doubly robust estimators without further structure. Subsequent work has developed doubly robust estimators under additional assumptions, including differentiability of the relevant functionals \cite{kennedy2017non}, monotonicity restrictions \cite{westling2020causal}, and smooth invertibility conditions \cite{diaz2023nonparametric}. The suitability of these assumptions is application-specific.

In this paper, we take a complementary approach. Rather than changing the target population (as in trimming) or targeting a policy parameter based on a pre-specified treatment transformation, we define a \emph{feasible dose--response curve} (FDRC) using a covariate-specific feasible-assignment rule. For each attempted dose, the rule assigns that dose when it is sufficiently supported among individuals with similar covariate histories, and otherwise assigns the nearest adequately supported dose. FDRC is the mean counterfactual outcome under this covariate-adaptive intervention assignment. It therefore preserves the original dose-response relationship where overlap is adequate, while remaining identifiable in regions with limited empirical support.

Our specific contributions are:
\begin{enumerate}
    \item \textbf{Diagnosing practical positivity violations for continuous interventions.}
    We develop a framework for assessing support under continuous treatment by defining covariate-specific feasible regions via highest-density regions of the conditional treatment distribution. Building on this definition, we introduce the \emph{non-overlap ratio}, a population-level diagnostic that summarizes the extent of support limitations across confounder strata and quantifies the practical severity of positivity violations.

    \item \textbf{Defining the most feasible intervention and the feasible dose--response curve.}
    We propose a feasible-intervention framework that maps infeasible target doses to nearby feasible values supported by the data. This induces a new estimand: feasible dose--response curve that remains close to the original CDRC while preserving interpretability and improving estimation under limited overlap.
\end{enumerate}

Overall, the proposed framework provides a practical and scientifically coherent approach for causal inference with continuous interventions when positivity is limited. We evaluate its performance in simulations and demonstrate its utility in a real-data application, showing improved stability and reduced reliance on unsupported extrapolation while retaining a target parameter closely aligned with the original dose--response question.

The remainder of the paper is organized as follows. Section~\ref{sec:motivation_example} presents the motivating data example. Section~\ref{sec:violation} introduces the support framework and the non-overlap ratio. Section~\ref{sec:mtp} defines the proposed feasible dose–response curve (FDRC) estimand and describes its identification, estimation strategy, and implementation algorithm. Section~\ref{sec:simulation} reports simulation results, and Section~\ref{sec:analysis} presents the real-data application. Section~\ref{sec:doscission} concludes with a discussion of implications and future directions. Section~\ref{sec:software} illustrates the practical implementation in the \texttt{CICI} package. 

\section{Motivation Example}\label{sec:motivation_example}

We motivate the problem using pharmacoepidemiological data from the CHAPAS-3 trial, which enrolled HIV-positive children in Zambia and Uganda \cite{Mulenga:2016, Abongomera:2017}. Our analysis focuses on children receiving efavirenz as part of combination antiretroviral therapy, together with lamivudine and either stavudine, zidovudine, or abacavir. Although dosing is weight-based, efavirenz \textit{concentration} exposure varies substantially across children due
to heterogeneity in metabolism---in particular, single nucleotide polymorphisms in the CYP2B6 gene encoding the key metabolizing enzyme---as well as adherence and other clinical factors. Because antiviral activity depends on exposure, subtherapeutic concentrations may lead to viral failure and subsequent regimen change. This motivates the scientific question: how does the probability of viral failure vary across efavirenz concentration levels?

Let EFV$_t$ denote the plasma efavirenz concentration (mg/L) at 12 hours after dosing at visit $t$, and let VL$_t$ indicate viral failure (viral load $>100$ copies/mL). A standard causal target is the concentration--response curve
\[
a \;\mapsto\; P\!\left(\mathrm{VL}_t^{\mathrm{EFV}=a}=1\right),
\]
defined over a clinically relevant range, e.g. $a \in (0,6)$. Informally, this estimand targets the question at which concentration levels in the blood the drug is still effective, and safe. Identification of this curve at a given level $a$ requires a positivity assumption relative to the estimand: within strata of covariates, the conditional treatment density must be bounded away from zero on the region where the curve is evaluated.

In pharmacological settings, this requirement can be difficult to satisfy because plasma concentration is not directly manipulable in the way an idealized intervention $\mathrm{EFV}_t=a$ suggests. Concentration is a downstream biomarker generated by a pharmacokinetic process linking dosing to absorption, distribution, metabolism, and elimination, and thus depends on individual characteristics and behaviors that are only partially controllable. For example, children with slow efavirenz clearance may be unable to achieve very low EFV$_t$ values even under clinically acceptable dose reductions, whereas fast metabolizers may rarely achieve very high concentrations at standard dosing. Weight- and age-based dosing guidelines further restrict the dose range available for a given child, and additional clinical constraints (e.g., toxicity concerns, comorbidities, drug--drug interactions, and regimen requirements) limit feasible dosing modifications. Adherence adds further structure: missed or delayed doses can yield low measured concentrations, but such values may occur primarily under specific adherence patterns and covariate profiles, making them sparse in the observed data.

When target concentration levels are weakly supported for a nontrivial subset of children, two considerations arise:
\begin{enumerate}
    \item \textbf{Statistical consideration.} Estimation of causal effects at those levels can be sensitive to model specification and finite-sample variability, because estimation may depend more heavily on extrapolation and/or unstable weighting when estimated treatment densities are very small.
    \item \textbf{Clinical/decision consideration.} Interpreting such targets as actionable interventions may be less straightforward, since some concentration levels may be difficult to achieve for many children under realistic clinical management.
\end{enumerate}
These considerations motivate prioritizing intervention values with adequate empirical support and considering estimands and methods that explicitly account for practical attainability.

In our application (Section~\ref{sec:analysis}), empirical support for EFV$_t$ is concentrated in the mid-range (approximately $1$–$3.5$ mg/L), with substantially weaker support near $0$ mg/L and above about $3.5$ mg/L, indicating limited overlap for target intervention values at the extremes (Figure~\ref{fig:ana1}). Estimation of $P(\mathrm{VL}_t^{\mathrm{EFV}=a}=1)$ at such extremes may therefore rely more heavily on extrapolation, and the estimated feasible dose-response curve departs from the standard concentration–response curve primarily in these regions, while remaining similar where support is adequate (Figure~\ref{fig:ana2}–\ref{fig:ana3}).

\section{Positivity and Diagnostics}\label{sec:violation}
\subsection{Notation}

Let \(O=(\vb L,A,Y)\in\mathcal O:=\mathcal L\times\mathcal A\times\mathcal Y\) denote one observed data unit, where
\begin{itemize}
    \item \(Y\in\mathcal Y\subseteq\mathbb R\) is the outcome,
    \item \(A\in\mathcal A\subseteq\mathbb R\) is the continuous treatment (or exposure),
    \item \(\vb L=(L_1,\dots,L_q)\in\mathcal L\subseteq\mathbb R^q\) is a vector of baseline covariates sufficient to control confounding.
\end{itemize}
We assume the temporal ordering
\[
L_1 \rightarrow L_2 \rightarrow \cdots \rightarrow L_q \rightarrow A \rightarrow Y,
\]
where each \(L_j\) is measured pre-treatment.

The observed sample is
\[
O_1,\dots,O_n \overset{\mathrm{iid}}{\sim} P_0,
\]
where \(P_0\) is the true data-generating distribution. Uppercase letters denote random variables and lowercase letters their realizations.

For each \(a\in\mathcal A\), let \(Y^a\) be the potential outcome \cite{rubin1974estimating} under the intervention that sets treatment to \(a\). The causal dose--response curve is
\begin{equation}
\label{eqn:cdrc}
m_0(a) := \E_0\!\left[Y^a\right], \qquad a\in\mathcal A.
\end{equation}

When densities exist (with respect to appropriate dominating measures), we use:
\begin{itemize}
    \item \(P_{0,A}\) and density \(f_{0,A}\) for the marginal law of \(A\),
    \item \(P_{0,\vb L}\) and density \(f_{0,\vb L}\) for the marginal law of \(\vb L\),
    \item \(P_{0,A\mid \vb L=\vb l}\) and density \(f_{0,A\mid \vb L}(a\mid \vb l)\) for the conditional law of \(A\mid \vb L=\vb l\).
\end{itemize}
Here and throughout, \(P\) denotes a generic distribution in the statistical model \(\mathcal M\), and \(P_n\) denotes the empirical measure.

\subsection{Determination of Positivity Violations}\label{sec:vio_det}

Identification of the causal dose--response curve in \eqref{eqn:cdrc} relies on the standard assumptions of consistency, exchangeability, and positivity. A common positivity condition \citep{petersen2012diagnosing} is
\begin{equation}\label{eqn:pos}
    \inf_{a \in \mathcal{A}} f_{0,A\mid \vb{L}}(a \mid \vb{L}) > 0, \quad \text{a.e.}
\end{equation}
where ``a.e.'' is with respect to \(P_{0,\vb L}\). Equivalently, for \(P_{0,\vb L}\)-almost every \(\vb l\), all treatment levels \(a\in\mathcal A\) relevant to \eqref{eqn:cdrc} have strictly positive conditional density under \(A\mid \vb L=\vb l\).

While \eqref{eqn:pos} is a population-level condition, empirical estimation depends on finite-sample support. In practice, some intervention levels may be weakly represented within certain covariate strata, which can reduce precision and increase reliance on extrapolation (see Appendix~\ref{apd:pos}). Thus, a practical diagnostic of support is needed.

A natural starting point is to classify support using a density threshold \(c>0\), as considered in \citet{Schomaker:2024}. For fixed \((a,\vb l)\), define
\begin{equation}
\left\{
\begin{aligned}
    &\text{supported}, && \text{if } f_{0,A\mid \vb{L}}(a \mid \vb{l}) > c,\\
    &\text{not supported}, && \text{if } f_{0,A\mid \vb{L}}(a \mid \vb{l}) \le c.
\end{aligned}
\right.
\end{equation}
This formalizes the idea that near-zero conditional density corresponds to weak support.

For binary treatments, this assessment is comparatively straightforward because conditional densities coincide with conditional probabilities. For continuous treatments, however, fixed cutoffs are less portable across settings: for each \(\vb l\),
\[
\int_{\mathcal A} f_{0,A\mid \vb L}(a\mid \vb l)\, da = 1,
\]
so the numerical scale of the density depends on the measurement scale and effective range of \(\mathcal A\). Consequently, universal thresholds (e.g., \(f_{0,A\mid \vb L}(a\mid \vb l)<0.001\)) are difficult to interpret consistently across studies.

Motivated by this limitation, we next introduce a diagnostic based on covariate-specific highest-density regions, providing a scale-aware characterization of practical positivity violations for continuous interventions.
\subsection{Proposal 1: Definition of Support and Diagnostic Using Highest Density Regions (HDRs)}\label{sec:diagnostic}

\subsubsection{Conditional support via highest density regions}

To characterize practical support for intervention values conditional on covariates, fix \textbf{support level} \(\alpha\in(0,1)\) and define a covariate-specific set \(\mathcal A_\alpha(\vb l)\subseteq\mathcal A\). Intuitively, \(\mathcal A_\alpha(\vb l)\) contains treatment values with relatively high conditional density under \(A\mid \vb L=\vb l\), chosen to contain \(\alpha\) of the conditional probability mass. We refer to \(a\in\mathcal A_\alpha(\vb l)\) as \emph{well supported} at level \(\alpha\); when no ambiguity arises, we write simply \emph{supported}. Values in \(\mathcal A\setminus\mathcal A_\alpha(\vb l)\) are \emph{poorly supported} (or \emph{unsupported}).

Formally, let \(f_{0,A\mid \vb L}(a\mid \vb l)\) be a version of the conditional density of \(A\mid \vb L=\vb l\) (with respect to a dominating measure on \(\mathcal A\)). Following the HDR construction of \citet{hyndman1996computing}, define a cutoff \(f_{0,\alpha}(\vb l)\) (Appendix~\ref{app:fa_quantile}) and set
\begin{equation}\label{eqn:falpha}
\mathcal A_\alpha(\vb l)
=
\{a\in\mathcal A:\ f_{0,A\mid \vb L}(a\mid \vb l)\geq f_{0,\alpha}(\vb l)\}
\end{equation}

such that
\[
P_{0,A\mid \vb L=\vb l}\!\big(\mathcal A_\alpha(\vb l)\big)=\alpha.
\]
Thus, for fixed \(\alpha\), intervention values outside \(\mathcal A_\alpha(\vb l)\) correspond to relatively low conditional density and are most prone to practical positivity violations.

If \(f_{0,A\mid \vb L}(a\mid \vb l)\) is constant on \(\mathcal A\) (e.g., \(A\mid \vb L=\vb l\) is uniform), the HDR is non-unique: any measurable subset of \(\mathcal A\) with conditional probability \(\alpha\) satisfies \eqref{eqn:falpha}. In that case, the well-supported/poorly-supported distinction is not informative.

\begin{figure}
    \centering
    \includegraphics[width=0.7\textwidth]{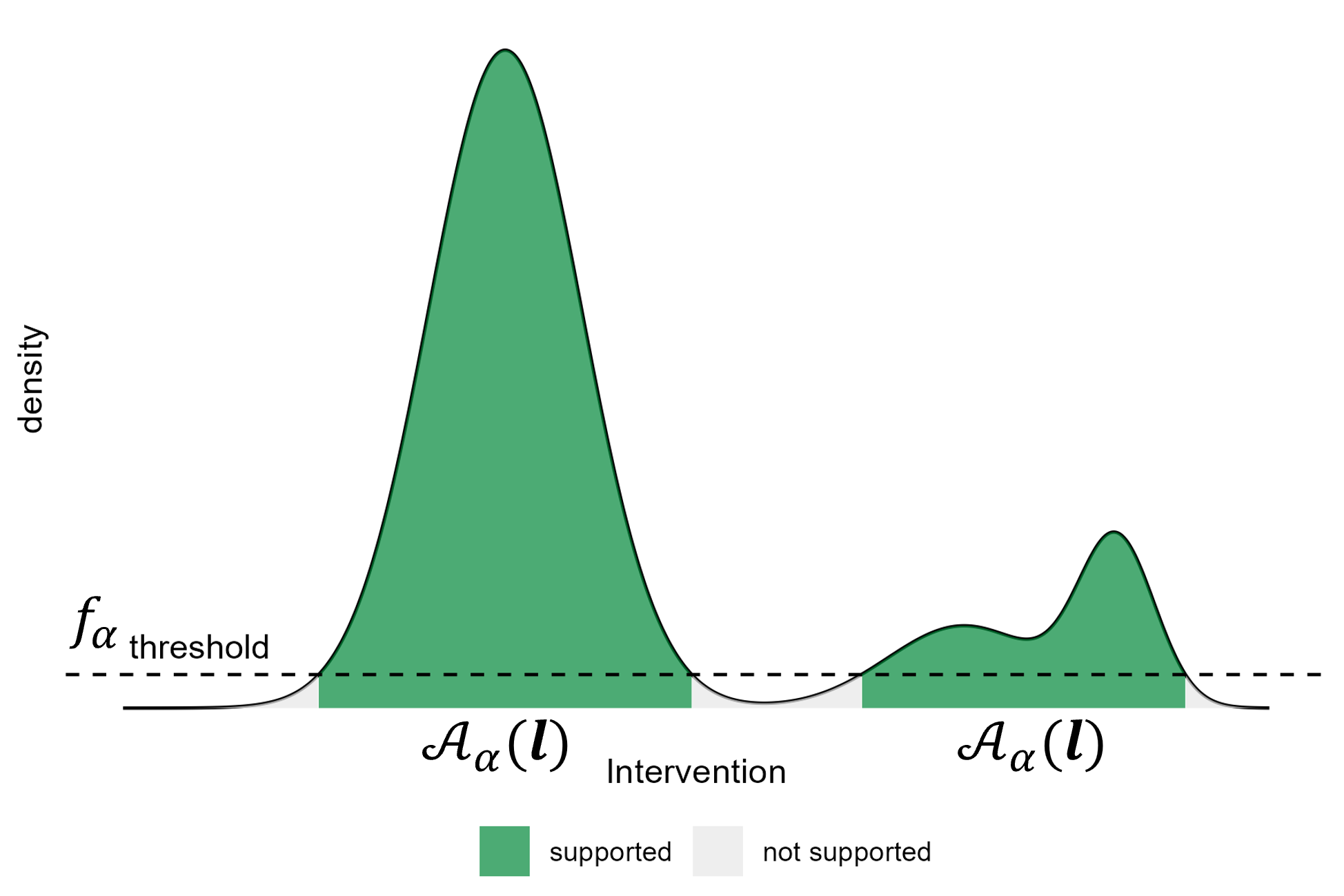}
    \caption{Illustration of the highest density regions (HDRs) used to define covariate-specific intervention support. For fixed \(\vb l\) and \(\alpha=0.95\), well-supported intervention values are those on the \(x\)-axis in \(\mathcal A_\alpha(\vb l)\). The green shaded region indicates probability mass of \(A\mid \vb L=\vb l\) inside \(\mathcal A_\alpha(\vb l)\); the grey region indicates mass outside.}
    \label{fig:high_density_region}
\end{figure}

Figure~\ref{fig:high_density_region} illustrates the construction for \(\alpha=0.95\): the green shaded area is the conditional probability mass inside \(\mathcal A_{0.95}(\vb l)\), and the grey area is the remaining mass outside.

\subsubsection{Diagnostic measure: the non-overlap ratio} 

To assess positivity in practice, we introduce the \textbf{non-overlap ratio}. This diagnostic summarizes, at the population level, the extent to which a candidate intervention value \(a\) falls outside the conditional support region defined by the HDRs. For a fixed \(\alpha\in(0,1)\), define \(\tau_{\alpha}:\mathcal A\to[0,1]\) by
\begin{equation}\label{eqn:nonoverlap}
    \tau_{\alpha}(a) := \int_{\mathcal L} \mathbbm{1}\{a \notin \mathcal{A}_{\alpha}(\vb{l})\}\, dP_{\vb{L}}(\vb{l})
    \;=\;
    \mathbb{P}\big(a\notin \mathcal A_{\alpha}(\vb{L})\big).
\end{equation}
Here, \(\tau_{\alpha}(a)\) represents the proportion of individuals whose covariate profiles \(\vb{L}\) render the intervention value \(a\) unsupported, i.e., lying outside the practical support region where positivity is regarded as plausible at level \(\alpha\).

The non-overlap ratio takes values in \([0,1]\):
\begin{itemize}
    \item Values near 0 indicate that \(a\) is well-supported across the population, with few or no positivity violations at the chosen support level.
    \item Values near 1 indicate that \(a\) is largely unsupported, reflecting severe positivity violations.
\end{itemize}

By aggregating individual-level support into a single summary, the non-overlap ratio provides a population-level diagnostic for evaluating the plausibility of the positivity assumption. Its practical use will be illustrated in the following sections.

\section{Estimands to Address Positivity Violations}
\label{sec:mtp}

Throughout this section, we view each estimand as a functional of an observed-data law \(P\) for \(O=(\vb L,A,Y)\), and denote its value under the true law \(P_0\). Specifically, for an estimand \(m^{\dagger}\) (\(\dagger\in\{\text{standard},\text{trim},\text{feasible}\}\)), we write \(m^{\dagger}(P)\) for the functional and \(m^{\dagger}_0:=m^{\dagger}(P_0)\) for the target under \(P_0\). Expectations and densities under \(P\) are denoted \(\mathbb E_P\) and \(f_P\), with \(\mathbb E_0\) and \(f_0\) corresponding to \(P_0\).

\subsection{Standard Causal Dose--Response Curve}\label{estmand:1}

\begin{boxwithhead}[Estimand 1 (CDRC)]
The standard causal dose--response curve is \(m^{\text{standard}}_0:\mathcal A\to\mathcal Y\) defined by
\[
m^{\text{standard}}_0(a)=\mathbb E_0\!\left[Y^{a}\right],\qquad a\in\mathcal A.
\]
Identification of \(m^{\text{standard}}_0(a)\) relies on the following standard conditions
for each \(a\in\mathcal A\):
\begin{enumerate}
\item \textbf{Consistency:} \(Y=Y^{a}\) when \(A=a\).
\item \textbf{Conditional exchangeability:} \(Y^{a}\perp A\mid \vb L\).
\item \textbf{Positivity:} \(f_{0,A\mid \vb L}(a\mid \vb l)>0\) for \(P_{0,\vb L}\)-almost every \(\vb l\).
\end{enumerate}
Under these conditions,
\begin{equation}\label{eqn:estimand1}
m^{\text{standard}}_0(a)
=\mathbb E_0\!\left[\mathbb E_0\!\left(Y\mid A=a,\vb L\right)\right].
\end{equation}
\end{boxwithhead}

The standard causal dose--response curve is conceptually appealing, but its identification and stable estimation hinge on the positivity assumption in the box above. In many applied continuous-intervention settings, this condition can be plausible for some intervention values and implausible for others. When \(f_{0,A\mid \vb L}(a\mid \vb l)\) is near zero in parts of the covariate space, estimation of \(m^{\text{standard}}_0(a)\) becomes fragile in the sense that estimation must rely on stronger modeling extrapolations across \(\vb l\), and common estimators can exhibit high variance and sensitivity to model misspecification because they are effectively learning about \(Y\) at \((A=a,\vb L=\vb l)\) from sparse data.

Importantly, these difficulties are not a consequence of continuity alone, but of limited empirical support for specific target intervention values within covariate strata. This is well illustrated by the CHAPAS-3 motivation (Section~\ref{sec:motivation_example}): EFV concentrations are downstream of dosing and pharmacokinetics and are therefore only partially controllable. As a result, some concentration levels are rarely observed for certain metabolic and clinical profiles, and the empirical support for EFV\(_t\) is concentrated in a mid-range, with substantially weaker support at the extremes (Section~\ref{sec:analysis}; Figure~\ref{fig:ana1}). In such regions, estimating the standard concentration--response curve at extreme \(a\) values is expected to be more dependent on extrapolation and correspondingly less stable, motivating the diagnostics and alternative estimands that explicitly restrict attention to well-supported interventions.

\subsection{Existing Estimand to Address Positivity Violations: Weighted Causal Dose--Response Curve}\label{sec:weighted}
\begin{boxwithhead}[Estimand 2 (Weighted CDRC)]
Schomaker et al.\ \cite{Schomaker:2024} propose the weighted CDRC, which is defined as
\begin{equation}\label{eqn:estimand_weighted}
    \begin{aligned}
        m^{\text{weighted}}_0(a)
        := &\, \mathbb{E}_0\!\left( Y^a \,\middle|\, f_0(a \mid \vb{L}) > c \right) 
            P_0\!\big(f_0(a \mid \vb{L}) > c\big) \\
        &+ \mathbb{E}_0\!\left( Y^a \, w_0(a \mid \vb{L}) \,\middle|\, f_0(a \mid \vb{L}) \leq c \right) 
            P_0\!\big(f_0(a \mid \vb{L}) \leq c\big),
            \qquad \forall a \in \mathcal{A},
    \end{aligned}
\end{equation}
where the weight function is
\begin{equation}
w_0(a, \vb{l}) =
\begin{cases}
    1, & \text{if } f_0(a \mid \vb{l}) > c, \\[6pt]
    \dfrac{f_0(a \mid \vb{l})}{f_0(a)}, & \text{if } f_0(a \mid \vb{l}) \leq c.
\end{cases}
\end{equation}

Under consistency and conditional exchangeability (as in Estimand~1), and assuming $f_0(a)>0$ (so that $w_0(a,\vb l)$ is well defined), $m^{\text{weighted}}_0(a)$ admits the identification formula
\begin{equation}
m^{\text{weighted}}_0(a)=\mathbb{E}_0\!\Big( \mathbb{E}_0\!\big(Y \mid A=a, \vb{L}\big)\, w_0(a, \vb{L}) \Big).
\end{equation}
\end{boxwithhead}

The weighted CDRC can be viewed as a modification of the standard CDRC that leaves the target unchanged on the well-supported subset $\{f_0(a\mid \vb L)>c\}$, while altering the contribution of covariate strata with low conditional density by introducing the weight $w_0(a,\vb L)$. This yields an estimand that is defined for all $a\in\mathcal A$ and is less sensitive to near-violations of positivity at treatment levels $a$ that are rarely observed within some covariate strata.

At the same time, the resulting target differs from the standard intervention estimand $m^{\text{standard}}_0(a)=\mathbb E_0(Y^a)$ whenever $P_0\{f_0(a\mid \vb L)\le c\}>0$. In those regions, $m^{\text{weighted}}_0(a)$ averages the conditional mean $\mathbb E_0(Y\mid A=a,\vb L)$ using weights that depend on the treatment density, so it no longer corresponds to the counterfactual mean under the single intervention that deterministically sets $A$ to $a$ for everyone. Rather, it is a density-weighted functional that coincides with the standard CDRC only where $f_0(a\mid \vb L)>c$. Consequently, the interpretation of the weighted curve depends on the choice of $c$ and on how one views the density-based reweighting in the low-support region.

More broadly, estimand modifications designed to address limited overlap trade direct correspondence to the standard ``set $A$ to $a$ for everyone'' target for improved stability when empirical support is sparse. The weighted CDRC is one principled example of this strategy: it provides a well-defined target across the full range of $a$, but its scientific meaning differs from $m^{\text{standard}}_0(a)$ whenever the modification is active. In applied work, it is therefore useful to report the weighted CDRC alongside overlap diagnostics and to be explicit about the implied target when $f_0(a\mid \vb L)$ falls below the cutoff.

\subsection{Novel Estimands to Address Positivity Violations: Feasible Dose–Response Curve (FDRC)}
\label{sec:feasible_estimand}

To mitigate the impact of positivity violations, we propose a causal estimand that restricts interventions to covariate-specific regions of the treatment space that are well supported under the data-generating mechanism. The key idea is to replace a target intervention level $a$ by a covariate-adaptive feasible value whenever $a$ falls outside the supported region for a given covariate profile, thereby reducing reliance on extrapolation while retaining a population-level causal target.

\begin{equation}
d_{0,\alpha}(a,\vb l)
=
\begin{cases}
a, & \text{if } a\in \mathcal A_\alpha(\vb l;P_0),\\[6pt]
\Pi_{\mathcal A_\alpha(\vb l;P_0)}(a), & \text{if } a\notin \mathcal A_\alpha(\vb l;P_0),
\end{cases}
\label{eqn:data_adaptive}
\end{equation}
where $\Pi_{\mathcal A_\alpha(\vb l;P_0)}(a)$ is the closest supported value to $a$:
\begin{equation}
\Pi_{\mathcal A_\alpha(\vb l;P_0)}(a)
=
\arg\min_{u\in\mathcal A_\alpha(\vb l;P_0)} |u-a|.
\label{eqn:most_feasible}
\end{equation}

Assume that for $P_0$-almost every $\vb l$, the set $\mathcal A_\alpha(\vb l;P_0)$ is nonempty and closed (for example, when $\mathcal A$ is compact). Then the minimum in \eqref{eqn:most_feasible} exists, and if multiple values are equally close, we apply a fixed tie-breaking rule (e.g., choose the smallest minimizer). Intuitively, this intervention rule keeps the target dose $a$ when it is supported for covariate profile $\vb l$, and otherwise replaces it with the nearest supported dose. For each fixed $a$, the map $\vb l\mapsto d_{0,\alpha}(a,\vb l)$ is therefore an individualized deterministic intervention rule.

\begin{figure*}
    \centering
    \subfloat[Targets outside the HDRs are reassigned to the closest feasible value within the HDRs. The green shaded area represents the highest density region (well-supported interventions), the grey area indicates regions with little or no support, and the dashed line marks the density threshold $f_\alpha$. Red arrows denote target interventions: Target~1 lies outside the HDR and is reassigned to the nearest feasible value (Most Feasible Intervention~1), while Target~2 lies within the HDR and is retained unchanged (Most Feasible Intervention~2).]{\includegraphics[width=0.49\textwidth]{HDR.pdf}\label{fig:feasible_intervention1}}
    \hspace{0.2cm}
    \subfloat[Targets inside the HDRs remain unchanged, while unsupported targets are shifted adaptively. Red circles denote target interventions, green circles represent their most feasible reassigned values, and black squares indicate observed data points across strata of the confounders. Green shaded areas indicate HDRs, while grey areas mark unsupported regions. This illustrates how infeasible interventions are reassigned adaptively to the closest feasible value, ensuring that interventions are only minimally adjusted.]{\includegraphics[width=0.49\textwidth]{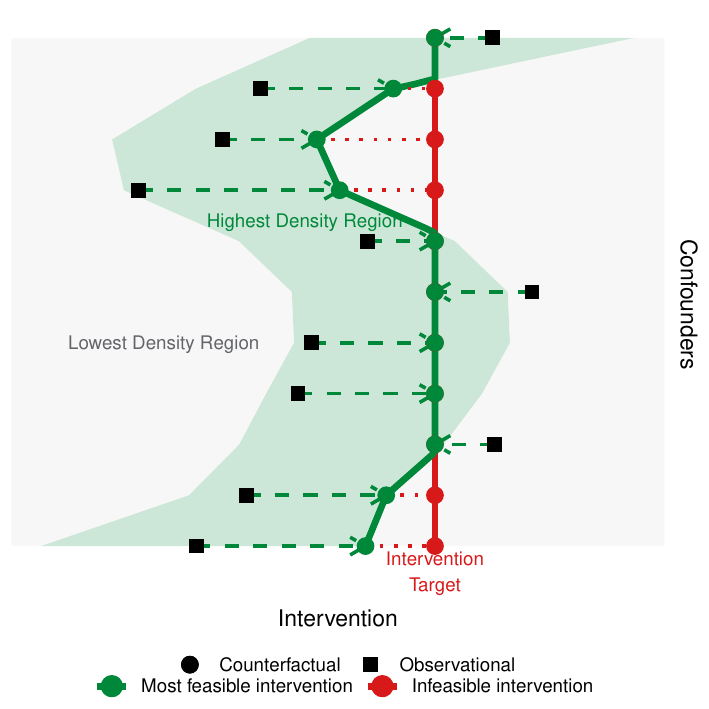}\label{fig:feasible_intervention2}}
    
    \caption{Illustration of the most feasible intervention defined in Equations~(\ref{eqn:data_adaptive}) and (\ref{eqn:most_feasible}). 
    Panel (a) shows the conditional density of treatment values given covariates, with HDRs (green) representing well-supported regions and grey areas denoting poorly supported regions. Target interventions (red arrows) are either retained if they fall inside HDRs or reassigned to the nearest feasible value within HDRs (green arrows). 
    Panel (b) extends this principle across confounder strata, where targets (red circles) are reassigned to the closest supported value (green circles) whenever they fall outside HDRs, while feasible targets remain unchanged. Together, the panels illustrate the principle of the most feasible intervention: retain the target when supported, and otherwise make the smallest necessary adjustment to ensure feasibility.}
    \label{fig:feasible_intervention}
\end{figure*}

An illustration is provided in Figure~\ref{fig:feasible_intervention1}. The horizontal axis shows intervention values and the vertical axis their estimated conditional density. Green shaded areas mark the highest density regions (HDRs), where the conditional density exceeds the threshold $f_\alpha$ (dashed line) and interventions are well supported; grey areas indicate poorly supported regions. The red arrows denote hypothetical targets: Target~1 lies outside the HDR, while Target~2 lies inside. Under the proposed rule, unsupported targets are reassigned to the nearest feasible value within the HDRs (green arrows). Thus, Target~1 shifts to the closest HDR point, while Target~2 remains unchanged. This illustrates the principle of the most feasible intervention: retain the target if supported, otherwise reassign it to the nearest supported value.

\begin{boxwithhead}[Estimand 3 (Feasible Dose-Response Curve)]
Fix \(\alpha\in(0,1)\). For each target level \(a\in\mathcal A\), let \(d_{0,\alpha}(a,\vb l)\) be the most feasible intervention operator defined in \eqref{eqn:data_adaptive}--\eqref{eqn:most_feasible}, so that the intervention assigns \(A\leftarrow d_{0,\alpha}(a,\vb L)\) and \(d_{0,\alpha}(a,\vb l)\in \mathcal A_\alpha(\vb l;P_0)\) for \(P_{0,\vb L}\)-almost every \(\vb l\). Let \(Y^{d_{0,\alpha}(a,\cdot)}\) denote the corresponding counterfactual outcome. The proposed feasible dose--response curve is
\begin{equation}
m_0^{\text{feasible}}(a)=\mathbb E_{0}\!\left\{Y^{d_{0,\alpha}(a,\cdot)}\right\},\qquad a\in\mathcal A.
\label{eqn:estimand_feasible}
\end{equation}

\smallskip
\noindent\textbf{Identification assumptions.} For each \(a\in\mathcal A\), assume:
\begin{enumerate}[label=\arabic*)]
\item \textbf{Consistency:} \(Y=Y^{a}\) when \(A=a\) for all \(a\in\mathcal A\); in particular, \(Y=Y^{d_{0,\alpha}(a,\cdot)}\) when \(A=d_{0,\alpha}(a,\vb L)\).
\item \textbf{Conditional exchangeability:} \(Y^{a}\perp A\mid \vb L\) for all \(a\in\mathcal A\).
\item \textbf{Well-definedness:} for \(P_{0,\vb L}\)-almost every \(\vb l\), \(\mathcal A_\alpha(\vb l;P_0)\) is nonempty (and closed), so that the projection in \eqref{eqn:most_feasible} exists; if the minimizer is not unique, ties are resolved by a deterministic measurable rule so that \((a,\vb l)\mapsto d_{0,\alpha}(a,\vb l)\) is measurable.
\end{enumerate}

\smallskip
\noindent Under 1)--3)\footnote{A formal derivation of this identification result is provided in Appendix~\ref{apd:proof_fdrc_identification}.
},
\[
m_0^{\text{feasible}}(a)
=
\mathbb E_{0}\!\left[
\mathbb E_{0}\!\left\{Y \mid A=d_{0,\alpha}(a,\vb L),\,\vb L\right\}
\right].
\]
\noindent 
Unlike the standard CDRC, identification does not require positivity at the target level \(a\) for all covariate strata; instead, by construction \(d_{0,\alpha}(a,\vb l)\in\mathcal A_\alpha(\vb l;P_0)\), so the estimand only intervenes on treatment levels that are well supported within each \(\vb l\)-stratum.
\end{boxwithhead}

\paragraph{Relation to dynamic interventions.}
The map \((a,\vb l)\mapsto d_{\alpha}(a,\vb l)\) is an individualized (covariate-dependent) intervention operator. For each fixed \(a\), the induced map \(\vb l\mapsto d_{\alpha}(a,\vb l)\) can be viewed as a (single-time) dynamic treatment rule. A key distinction from the dynamic regimes most commonly studied in causal inference is that our rule is defined through a covariate-specific supported set \(\mathcal A_\alpha(\vb l;P_0)\), which is a functional of the conditional exposure distribution under \(P_0\). Thus the covariate dependence is introduced for feasibility: for each \((a,\vb l)\), the rule coincides with the static intervention “set \(A\) to \(a\)” when \(a\in\mathcal A_\alpha(\vb l;P_0)\), and otherwise maps \(a\) to a nearby supported value.

\paragraph{Comparison of Estimand 2 and Estimand 3} 
Both the weighted CDRC and the FDRC aim to address positivity violations by restricting attention to regions with sufficient support. 
However, they adapt the estimand in fundamentally different ways. 
The weighted CDRC of Schomaker et al.\ \cite{Schomaker:2024} applies a fixed cutoff \(c\) to the conditional density and redefines the target through projection weights. 
This ensures identifiability without extra assumptions, but it alters the causal parameter: in regions of support, the curve coincides with the standard CDRC, while in unsupported strata it relies on weighted associational components. 
As Schomaker et al.\ emphasize, this means the weighted curve sticks to the causal question where feasible but falls back on associations where necessary, yielding a compromise between interpretability and stability. 
By contrast, the FDRC employs a unit-specific threshold and modifies intervention assignments explicitly by reassigning unsupported values to the closest feasible ones. 
This produces a transparent intervention rule targeting the well-defined quantity \(E(Y^{d_{0,\alpha}(a,\cdot)})\). 
In short, the weighted CDRC trades some causal interpretability for broader identifiability, whereas the FDRC preserves interpretability while adapting directly to the actual support.

\paragraph{Interpretation}

Figure~\ref{fig:feasible_intervention2} illustrates how the most feasible intervention strategy operates at the individual level, conditional on confounders.
The vertical axis represents confounder values, and the horizontal axis represents the intervention.
The shaded green band marks the highest density regions, where intervention values are well supported by the data, while the surrounding grey area corresponds to lowest density regions with little or no support.
The red vertical line indicates the oracle intervention target, which may not be feasible for all individuals.
Units whose oracle intervention lies outside the highest density regions (shown as red circles) cannot be assigned this target directly.
Instead, the proposed rule reassigns them to the nearest feasible value within the highest density regions, shown by the green circles and arrows.
By contrast, units whose oracle target already falls within the highest density regions (green circles on the red line) retain their original assignment.
Black squares represent the observed data (intervention--confounder pairs), from which the highest density regions are estimated.
Overall, the figure demonstrates that the most feasible intervention rule systematically reassigns unsupported targets to the closest feasible values, ensuring that every individual receives an intervention assignment compatible with the data distribution.

Violations of the positivity assumption are not merely theoretical concerns but also reflect real-world scenarios where certain interventions are impractical or unrealistic for specific subpopulations. For example, consider drug concentration levels across patients, as discussed in Section \ref{sec:analysis}. Patients with a fast metabolism, who metabolize drugs rapidly, naturally exhibit low drug concentrations even when administered standard doses. For these patients, a high drug concentration target would be unrealistic because it would require doses far beyond what is commonly observed or safely prescribed. While theoretically possible, achieving such concentrations would involve dosing regimens that are rarely practiced due to clinical and practical concerns. This leads to a situation where data corresponding to such high concentrations is either extremely sparse or non-existent, resulting in biased causal inferences and making conclusions drawn from such targets unreliable.

In this setting, the feasible dose--response curve (FDRC) answers a simple, real-world question: for each concentration level we \emph{aim for}, what outcome would we expect on average if children who can realistically reach that level are set to it, while children for whom that level is not realistically attainable are instead moved to the closest concentration they can realistically attain? In other words, the FDRC describes what would happen if we target a given concentration but allow individual “fallback” adjustments when the target is out of reach. It therefore matches the usual dose--response curve where targets are broadly attainable, and it differs mainly at extreme targets that are unrealistic for some children.

The proposed feasible dose--response curve addresses these issues by adapting the intervention target to remain within regions with adequate empirical support. Instead of imposing a target that is weakly supported by the observed data, this approach modifies the intervention according to what is practically achievable and observable in the data. This has two key benefits: first, it can improve robustness and stability of estimation in the presence of sparsity and practical positivity violations (relative to unsupported targets), and second, it ensures that the causal effects being estimated are relevant and actionable in practice.

Unlike static interventions, which risk oversimplifying causal relationships by imposing uniform targets, the most feasible intervention approach offers a more nuanced and reasonable strategy. It acknowledges the inherent variability in real-world data and adapts the intervention accordingly, leading to more robust, reliable, and meaningful causal inferences that better reflect both the data and the practical realities of clinical and observational settings.

\subsection{Trimmed dose–response curve}\label{estmand:3}


\begin{boxwithhead}[Estimand 4 (Trimmed Dose--Response Curve)]
Fix \(\alpha\in(0,1)\). For each target level \(a\in\mathcal A\), define
\begin{equation}
\label{eqn:estimand_trim}
m_{0}^{\mathrm{trim}}(a)
:=
\mathbb E_{0}\!\left[Y^{a}\mid a\in\mathcal A_{\alpha}(\vb L;P_{0})\right],
\qquad a\in\mathcal A,
\end{equation}
whenever \(P_{0}\!\big(a\in\mathcal A_{\alpha}(\vb L;P_{0})\big)>0\), equivalently \(\tau_{0,\alpha}(a)<1\), where \(\tau_{0,\alpha}\) the non-overlap ratio is defined in \eqref{eqn:nonoverlap}.

\smallskip
\noindent\textbf{Identification assumptions.} For each \(a\in\mathcal A\), assume:
\begin{enumerate}[label=\arabic*)]
\item \textbf{Consistency:} \(Y=Y^{a}\) when \(A=a\).
\item \textbf{Conditional exchangeability:} \(Y^{a}\perp A\mid \vb L\).
\item \textbf{Well-defined retained stratum:} \(P_{0}\!\big(a\in\mathcal A_{\alpha}(\vb L;P_{0})\big)>0\) (equivalently \(\tau_{0,\alpha}(a)<1\)).
\end{enumerate}

\smallskip
\noindent Under 1)--3), this estimand is identified by
\[
m_{0}^{\mathrm{trim}}(a)
=
\frac{
\int
\mathbb E_{0}\!\left(Y\mid A=a,\vb L=\vb l\right)
\mathbbm 1\!\{a\in\mathcal A_{\alpha}(\vb l;P_{0})\}
\,dP_{0,\vb L}(\vb l)
}{
P_{0}\!\big(a\in\mathcal A_{\alpha}(\vb L;P_{0})\big)
}.
\]
Equivalently,
\[
m_{0}^{\mathrm{trim}}(a)
=
\int
\mathbb E_{0}\!\left(Y\mid A=a,\vb L=\vb l\right)\,
t_{0,\alpha}(a,\vb l)\,
dP_{0,\vb L}(\vb l),
\]
where
\[
t_{0,\alpha}(a,\vb l)
=
\frac{\mathbbm 1\!\{a\in\mathcal A_{\alpha}(\vb l;P_{0})\}}
{1-\tau_{0,\alpha}(a)}.
\]

\smallskip
\noindent Similar to FDRC, identification does not require positivity at level \(a\) across all covariate strata; instead, by conditioning on \(a\in\mathcal A_{\alpha}(\vb L;P_{0})\), the estimand targets strata where \(a\) is supported.
\end{boxwithhead}

\noindent\textbf{Comparison:}
Relative to conventional trimming estimands, this formulation replaces the fixed density threshold \(c\) with a support restriction defined by the conditional HDR at level \(\alpha\). This makes the notion of ``sufficient support'' explicit on the treatment scale through membership in \(\mathcal A_\alpha(\vb l;P_0)\), rather than through an arbitrary cutoff on \(f_{0,A\mid \vb L}(a\mid \vb l)\).

As with any trimming estimand, the target parameter is conditional on a covariate-dependent retained set---here, strata with \(a\in\mathcal A_\alpha(\vb L;P_0)\). This can improve statistical stability by reducing extrapolation to weak-support regions, but it also changes the causal target and may limit transportability to the full population.


\subsection{Estimation}\label{sec:estimation}

\subsubsection{Estimation Algorithm}
To estimate the standard CDRC in \eqref{eqn:estimand1}, a parametric g-formula can be applied directly for single time-point interventions. Substitution (plug-in) estimators can likewise be constructed for the population-level estimands in \eqref{eqn:estimand_feasible} and \eqref{eqn:estimand_trim}; the full procedure is given in Algorithm~\ref{alg:feasible}.

The complete CDRC and FDRC estimation workflow is implemented in the \texttt{CICI} package \cite{Schomaker:2024,manualcici}, and illustrated in the Software Implementation section (Section~\ref{sec:software}). The software follows the same steps as Algorithm~\ref{alg:feasible}.

\begin{algorithm}
\caption{Plug-in estimation of \(\hat m_n^{\mathrm{standard}}\), \(\hat m_{n,\alpha}^{\mathrm{feasible}}\), \(\hat m_{n,\alpha}^{\mathrm{trim}}\), and \(\hat\tau_{n,\alpha}\) at target values}
\label{alg:feasible}
\begin{algorithmic}[1]

\State \textbf{Inputs:}
\(\{(\vb L_j,A_j,Y_j)\}_{j=1}^n\);
target grid \(\overline{\vb a}^{\,\mathrm{tar}}=(a^{\mathrm{tar}}_1,\ldots,a^{\mathrm{tar}}_{m_{\mathrm{tar}}})^\top\subset\mathcal A\);
HDR grid \(\overline{\vb a}^{\,\mathrm{hdr}}=(a^{\mathrm{hdr}}_1,\ldots,a^{\mathrm{hdr}}_{m_{\mathrm{hdr}}})^\top\subset\mathcal A\),
assumed strictly increasing.
Let \((b_k)_{k=1}^{m_{\mathrm{hdr}}+1}\) be the induced cell boundaries defined by
\[
b_{1}:=a^{\mathrm{hdr}}_{1},\qquad
b_{k}:=\frac{a^{\mathrm{hdr}}_{k-1}+a^{\mathrm{hdr}}_{k}}{2}\ \ (k=2,\dots,m_{\mathrm{hdr}}),\qquad
b_{m_{\mathrm{hdr}}+1}:=a^{\mathrm{hdr}}_{m_{\mathrm{hdr}}},
\]
and let \(\{I_k\}_{k=1}^{m_{\mathrm{hdr}}}\) be the associated partition of
\([a^{\mathrm{hdr}}_1,a^{\mathrm{hdr}}_{m_{\mathrm{hdr}}}]\) given by
\[
I_k := [b_k,b_{k+1}) \ (k=1,\dots,m_{\mathrm{hdr}}-1), 
\qquad
I_{m_{\mathrm{hdr}}}:=[b_{m_{\mathrm{hdr}}},b_{m_{\mathrm{hdr}}+1}],
\]
with widths \(w_k:=\lambda(I_k)=b_{k+1}-b_k>0\).
Support level \(\alpha\in(0,1)\).
Fix a deterministic tie-breaking rule for projections \(\Pi_T(\cdot)\).

\State \textbf{Estimate treatment density.}
Fit \(\hat f_n(a\mid \vb l)\) (see Section \ref{sec:cde}), and evaluate:
\(\hat f_n(a^{\mathrm{hdr}}_k\mid \vb L_j)\) for all \(k,j\),
and \(\hat f_n(a^{\mathrm{tar}}_i\mid \vb L_j)\) for all \(i,j\).

\State \textbf{Construct subject-specific HDR cutoffs and supported sets.}
For each \(j=1,\ldots,n\):
\begin{enumerate}[label=(\alph*)]
\item Compute \(\hat p_{n,k}(\vb L_j)\) by \eqref{eq:grid_weights}, using \(w_k=\lambda(I_k)\).
\item Compute \(\hat f_{n,\alpha}(\vb L_j)\) by \eqref{eq:fa_quantile_hat}.
\item Define the estimated supported set on the HDR grid:
\[
\hat{\mathcal A}^{\,\mathrm{hdr}}_{n,\alpha}(\vb L_j)
=
\Big\{
a^{\mathrm{hdr}}_k\in\mathcal A:\ 
\hat f_n(a^{\mathrm{hdr}}_k\mid \vb L_j)\ge \hat f_{n,\alpha}(\vb L_j),
\ k=1,\ldots,m_{\mathrm{hdr}}
\Big\}.
\]
\end{enumerate}

\State \textbf{Estimate non-overlap ratio at target values.}
For each \(i=1,\ldots,m_{\mathrm{tar}}\),
\[
\hat\tau_{n,\alpha}(a^{\mathrm{tar}}_i)
=
\frac{1}{n}\sum_{j=1}^n
\mathbbm 1\!\left\{\hat f_n(a^{\mathrm{tar}}_i\mid \vb L_j)<\hat f_{n,\alpha}(\vb L_j)\right\}.
\]

\State \textbf{Estimate outcome regression.}
Fit \(\hat\mu_n(a,\vb l)\approx \E(Y\mid A=a,\vb L=\vb l)\).

\State \textbf{Compute plug-in curves at each target \(a^{\mathrm{tar}}_i\).}
For each \(i=1,\ldots,m_{\mathrm{tar}}\):
\begin{enumerate}[label=(\alph*)]
\item \emph{Standard CDRC:}
\[
\hat m_n^{\mathrm{standard}}(a^{\mathrm{tar}}_i)
=
\frac{1}{n}\sum_{j=1}^n \hat\mu_n(a^{\mathrm{tar}}_i,\vb L_j).
\]

\item \emph{Feasible FDRC:}
\[
\hat d_{n,\alpha}(a^{\mathrm{tar}}_i,\vb L_j)=
\begin{cases}
a^{\mathrm{tar}}_i, &
\hat f_n(a^{\mathrm{tar}}_i\mid \vb L_j)\ge \hat f_{n,\alpha}(\vb L_j),\\
\Pi_{\hat{\mathcal A}^{\,\mathrm{hdr}}_{n,\alpha}(\vb L_j)}(a^{\mathrm{tar}}_i), &
\hat f_n(a^{\mathrm{tar}}_i\mid \vb L_j)< \hat f_{n,\alpha}(\vb L_j),
\end{cases}
\]
\[
\hat m_{n,\alpha}^{\mathrm{feasible}}(a^{\mathrm{tar}}_i)
=
\frac{1}{n}\sum_{j=1}^n
\hat\mu_n\!\big(\hat d_{n,\alpha}(a^{\mathrm{tar}}_i,\vb L_j),\vb L_j\big).
\]

\item \emph{Trimmed curve:}
\[
\hat t_{n,\alpha}(a^{\mathrm{tar}}_i,\vb L_j)
=
\frac{
\mathbbm 1\!\left\{\hat f_n(a^{\mathrm{tar}}_i\mid \vb L_j)\ge \hat f_{n,\alpha}(\vb L_j)\right\}
}{
1-\hat\tau_{n,\alpha}(a^{\mathrm{tar}}_i)
},
\]
\[
\hat m_{n,\alpha}^{\mathrm{trim}}(a^{\mathrm{tar}}_i)
=
\frac{1}{n}\sum_{j=1}^n
\hat\mu_n(a^{\mathrm{tar}}_i,\vb L_j)\,
\hat t_{n,\alpha}(a^{\mathrm{tar}}_i,\vb L_j).
\]
\end{enumerate}

\State \textbf{(Optional) Uncertainty quantification.}
Use the nonparametric bootstrap for approximate compatibility intervals.

\end{algorithmic}
\end{algorithm}

\medskip
\noindent\textbf{Remarks:}
\begin{enumerate}[label=\arabic*)]
\item
The estimated conditional treatment density \(\hat f_n(a\mid \vb l)\) (Algorithm~\ref{alg:feasible}, Step~2) is used only to construct subject-specific cutoffs \(\hat f_{n,\alpha}(\vb l_j)\) and hence estimated supported sets
\[
\hat{\mathcal A}^{\,\mathrm{hdr}}_{n,\alpha}(\vb l_j)
=
\left\{a^{\mathrm{hdr}}_k\in\mathcal A:\hat f_n(a^{\mathrm{hdr}}_k\mid \vb l_j)\ge \hat f_{n,\alpha}(\vb l_j)\right\},
\]
as well as derived quantities \(\hat d_{n,\alpha}(a,\vb l_j)\), \(\hat t_{n,\alpha}(a,\vb l_j)\), and \(\mathbbm 1\{a\in\hat{\mathcal A}^{\,\mathrm{hdr}}_{n,\alpha}(\vb l_j)\}\) (Steps~6b--6c). It is not included as a covariate in the outcome regression \(\hat\mu_n(a,\vb l)\), except through these constructions.

\item
The HDR cutoff computation used in Step~3 is the grid-based sample analogue of the population quantile characterization in Appendix~\ref{app:fa_quantile}. Specifically, the population cutoff \(f_\alpha(\vb l)\) is defined as the \((1-\alpha)\)-quantile of the density-value variable \(Z_{\vb l}=f_0(A\mid \vb l)\) in \eqref{eq:fa_quantile_pop}, and the corresponding HDR is the closed upper level set in \eqref{eq:hdr_closed_def}. Algorithm~\ref{alg:feasible} implements this via the weighted discrete approximation \(\hat p_{n,k}(\vb l)\) in \eqref{eq:grid_weights}, the estimated quantile \(\hat f_{n,\alpha}(\vb l)\) in \eqref{eq:fa_quantile_hat}, and the estimated closed upper level set \(\hat{\mathcal A}^{\,\mathrm{hdr}}_{n,\alpha}(\vb l)\) in \eqref{eq:hdr_hat_closed_def}. The grid-cell widths \(w_k\) ensure proper normalization when the HDR grid is not equally spaced.

\item
A formal consistency result for the feasible estimator is given in Appendix~\ref{apd:proof_fdrc} and ~\ref{apd:proof_fdrc:main} . Under Assumptions (A1)--(A8) in Section~\ref{apd:proof_fdrc:assumptions}, Theorem~\ref{apd:proof_fdrc:thm} shows that for each fixed \(a\in\mathcal A\),
\[
\hat m^{\mathrm{feasible}}_{n,\alpha}(a)\xrightarrow{P} m^{\mathrm{feasible}}_{0,\alpha}(a).
\]
The argument combines (i) consistency of \(\hat f_n\) and \(\hat f_{n,\alpha}\), which stabilizes estimated support membership (Lemma~\ref{apd:proof_fdrc:lemma1}); (ii) convergence of the estimated feasible rule \(\hat d_{n,\alpha}(a,\vb l)\to d_{0,\alpha}(a,\vb l)\) (Lemma~\ref{apd:proof_fdrc:lemma_rule}); and (iii) consistency of \(\hat\mu_n\), which controls plug-in regression error.

\item
Step~2 accommodates multiple conditional-density estimators (e.g., parametric, binning, hazard-binning, HAL-based approaches). The hazard-binning construction and implementation details are provided in Section~\ref{apd:hazard}; the framework in Algorithm~\ref{alg:feasible} is unchanged across these options.

\item
Additional remarks on inference, bootstrap validity, first-stage nuisance estimation, and sample-splitting considerations are provided in Appendix~\ref{apd:estimation_additional_notes}.
\end{enumerate}

\subsubsection{Conditional Density Estimation}
\label{sec:cde}
A key step is estimation of the conditional treatment density \(f_{A\mid \vb L}(a\mid \vb l)\), which determines supported intervention regions and subject-specific feasibility cutoffs. Several estimators can be used, including parametric models, nonparametric kernel conditional density estimators \cite{hayfield2008nonparametric}, and flexible machine-learning approaches such as HAL-based density estimation (Haldensify) \cite{hejazi2022efficient, hejazi2022haldensify-joss, hejazi2022haldensify-rpkg}. 

In our implementation, we allow multiple density-estimation options within the same plug-in framework. One option is a hazard-binning estimator motivated by Haldensify and related to the pooled-hazard approach of \citet{diaz2011super}; its construction is described in Section~\ref{apd:hazard}. Regardless of the specific estimator used, the downstream steps in Algorithm~\ref{alg:feasible} (HDR construction, feasible mapping, trimming weights, and plug-in evaluation) remain the same.

\begin{remark}[Grid-based versus closed-form HDR construction]
\label{rem:cde_grid}
Some conditional density estimators are naturally implemented on a prespecified grid in \(a\).
In particular, grid-based methods such as Haldensify and the hazard-binning estimator
(Section~\ref{apd:hazard}) return \(\hat f_n(a\mid \vb l)\) through evaluations on a finite grid,
so we construct the HDR and the subject-specific cutoff using the HDR grid
\(\overline{\vb a}^{\,\mathrm{hdr}}\) together with its induced cell widths \(\{w_k\}\)
(Appendix~\ref{app:fa_quantile:sample}).

In contrast, for fully parametric conditional densities (e.g., Gaussian regression), the HDR can in principle be characterized directly as an upper level
set of \(\hat f_n(\cdot\mid \vb l)\) without discretization. Nevertheless, even in the
parametric case, the same grid-based construction is often convenient and numerically
stable, and it matches the pragmatic implementation used in Algorithm~\ref{alg:feasible}.
\end{remark}

\section{Simulation Studies}\label{sec:simulation}

In this section, we examine the proposed methods through three distinct simulation settings. These simulations are designed to provide insight into the performance of the methods under different conditions.

In Simulation 1, we explore how positivity violations exacerbate bias in estimation, particularly when the outcome models are misspecified. Simulation 2 examines the impact of positivity violations on estimation by comparing scenarios with and without positivity violations. Simulation 3 investigates the methods in a more complex setting, incorporating multiple confounders and intricate data-generating processes.

\subsection{Simulation Settings}
\subsubsection{Simulation 1: Correct model specification versus misspecification under positivity violations}

We consider a simple scenario with a single time point where both the intervention and the confounder are normal-distributed shown in Appendix \ref{apd:sim1}. The simulation study involves three variations that share the same distribution for the confounder and intervention but differ in their outcome functions.

In Simulation 1A, the true outcome function is linear. In Simulation 1B, the true outcome function is sinusoidal, and in Simulation 1C, it is logarithmic. These two non-linear outcome functions in Simulations 1B and 1C pose a greater challenge for estimation, particularly in regions where data is sparse, i.e. suffering from positivity violations, leading to potential estimation difficulties. In contrast, the linear outcome function in Simulation 1A is expected to be easier to estimate, even in the presence of data sparsity, as the extrapolation will work perfectly.

The primary objective of this simulation design is to evaluate and compare the bias and variability in estimating both the standard CDRC, the proposed CDRC with most feasible intervention, and also the trimming estimand under conditions with positivity violations. This comparison provides insights into the strengths of the proposed most feasible intervention, particularly when outcome functions can not be reliably extrapolated.

\subsubsection{Simulation 2: Correct model specification with positivity violations versus no positivity violations}

We consider a different scenario with a single time point where the confounder is binary shown in Appendix \ref{apd:sim2}. The simulation study involves two variations that share the same distribution for the confounder and but differ in their intervention distribution.

In both variations, the intervention distributions given the confounder follow a truncated normal distribution with the same mean value conditional on $L$, but differ in variance. Simulation 2A has a lower variance, leading to greater deviation compared to Simulation 2B. Specifically, for $L = 0$ and $L = 1$, the overlap between the distributions of $A|L$ is minimal, indicating a higher degree of positivity violations, as shown in Figure \ref{fig:sim2A} and \ref{fig:sim2B}. The outcome function is a linear combination of $A$ and $L$ with a small noise term, making it straightforward to estimate and suitable for extrapolation.

The two simulations are designed to compare the estimation of the standard CDRC and the new estimands under conditions of positivity violations (2A) and no positivity violations (2B). This comparison will highlight the performance differences between the three estimands in scenarios where the positivity assumption is either met or violated.

\subsubsection{Simulation 3: Simulation study in more complex setting inspired by real data}

In this scenario, we consider a complex data-generating process, detailed in Appendix \ref{apd:sim3}. This process is inspired by real data from the CHAPAS-3 study, an open-label, parallel-group, randomized trial \cite{Mulenga:2016}, which is further analyzed in Section \ref{sec:analysis} and follows the setup described in Schomaker et al. \cite{Schomaker:2024}.

The outcome variable is binary and exhibits a non-linear relationship with the continuous intervention variable, which is modeled using a truncated normal distribution. The simulation aims to validate the proposed approaches in complex scenarios involving intricate distributions and diverse confounders with real-world significance.

\subsection{Estimation and Evaluation}

We set the sample size to \(n = 1000\) for all simulation studies. The true value of each estimand \(m^{\dagger}_0\), where \(\dagger \in \{\text{standard}, \text{feasible}, \text{trim}\}\), is approximated by intervening on the data-generating process under the corresponding action using a large Monte Carlo sample of size \(N = 100{,}000\). For estimation (see Algorithm~\ref{alg:feasible} for details), we apply the parametric g-computation plug-in estimator in Simulations~1 and~2, and the HAL-based estimator in Simulation~3. The method for conditional density estimation varies by scenario: parametric linear regression (Simulation~1), nonparametric kernel methods (Simulation~2, using the R package \texttt{np}~\cite{hayfield2008nonparametric}), and hazard-binning (Simulation~3; see Appendix~\ref{apd:hazard}). We assess bias over \(R = 1000\) Monte Carlo replications, defining absolute bias at level \(a\) as
\[
\text{Absolute Bias}(a) = 
\dfrac{1}{R}\sum_{r=1}^R \left|\hat{m}^{\dagger}_r(a) - m^{\dagger}_0(a)\right|.
\]

\subsection{Results}

The results of the simulations are summarized in Figures \ref{res:sim1} to \ref{res:sim3}.

\subsubsection{Simulation 1}

\begin{figure*}
    \centering
    \subfloat[Conditional density distribution of $A|L$ with different support level and scatter plot of observed data from one single simulation run. \label{fig:sim1A}]{\includegraphics[width=0.45\textwidth]{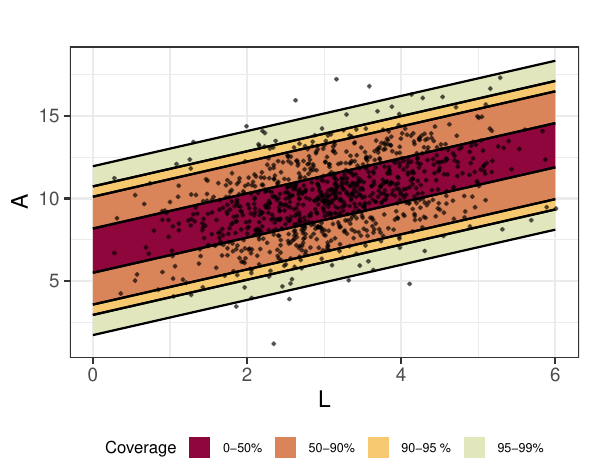}}
    \hspace{0.2cm}
    \subfloat[Non-overlap ratio: the conditional support with support level of $95\%$. \label{fig:sim1B}]{\includegraphics[width=0.45\textwidth]{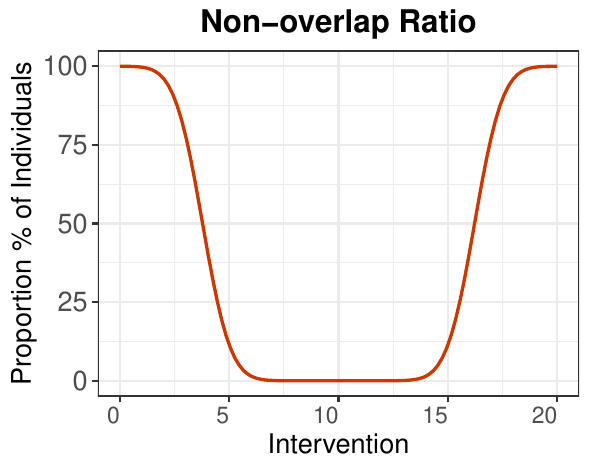}}
    
    \subfloat[Comparison of true and estimated values of \(m^\text{standard}\), \(m^\text{feasible}\), and \(m^\text{trimming}\) from Simulation 1. Separated by linear, logarithmic, and sinusoidal outcome functions across three simulation settings.\label{fig:sim1C}]{\includegraphics[width=0.95\textwidth]{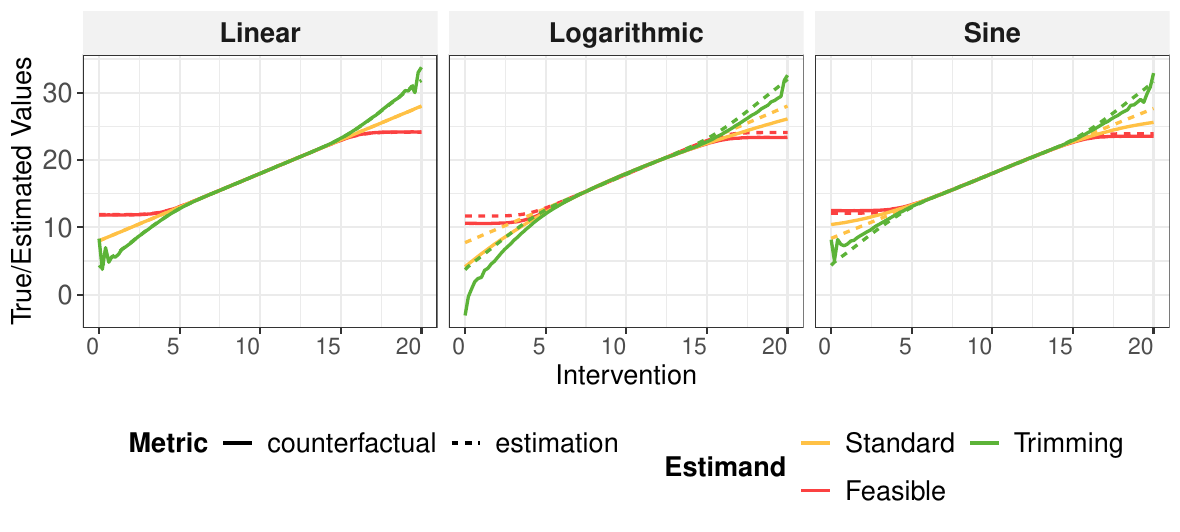}}
    
    \subfloat[Absolute bias of the estimation of \(m^\text{standard}\), \(m^\text{feasible}\), and \(m^\text{trimming}\) from Simulation 1. Separated by linear, logarithmic, and sinusoidal outcome functions across three simulation settings.\label{fig:sim1D}]{\includegraphics[width=0.95\textwidth]{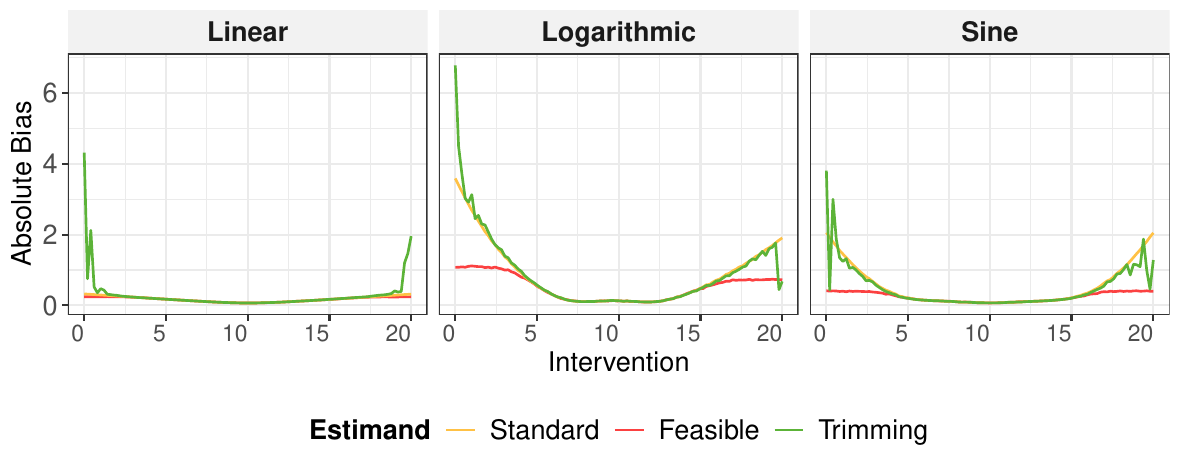}}
    \caption{Results of Simulation 1}
    \label{res:sim1}
\end{figure*}

Figure \ref{fig:sim1A} illustrates the estimated conditional distribution and support levels at $50\%$, $90\%$, $95\%$, and $99\%$, overlaid with a scatter plot of observed data from a single simulation run. Areas outside these regions highlight rare combinations of $L$ and $Y$ in observation, reflecting data sparsity. Using the diagnostic approach from Section \ref{sec:vio_det}, we identify positivity violations, as shown in Figure \ref{fig:sim1B}.

Figure \ref{fig:sim1B} presents the non-overlap ratio, a diagnostic metric quantifying the proportion of the population for which specific interventions are infeasible at a $95\%$ support level. This ratio, ranging from $0$ to $1$, reveals a clear pattern: in the central region, nearly all interventions have a non-overlap ratio of zero, indicating good feasibility. In contrast, the ratio rises sharply at the extremes, indicating that extremely small or large interventions are largely infeasible.

Together, Figures \ref{fig:sim1A} and \ref{fig:sim1B} show that certain interventions are rare or infeasible for subsets of the population. Even theoretically feasible, interventions may lack sufficient data for accurate estimation, underscoring the importance of our proposed feasible intervention framework.

Figure \ref{fig:sim1C} examines the impact of non-overlap and model misspecification on the estimation of three estimands under linear, logarithmic, and sinusoidal outcome functions. In regions where the non-overlap ratio is near zero, counterfactual estimates for the three estimands are nearly identical across all outcome functions, as intended by the simulation design. At the extremes, however, the estimands diverge. The FDRC establishes practical thresholds, avoiding unrealistic extrapolation into sparsely supported regions. The trimming estimand, by contrast, restricts the population to those for whom the intervention is feasible.

For linear outcome functions (Figure \ref{fig:sim1C}, left), extrapolation is highly accurate, resulting in minimal absolute bias across all estimands (Figure \ref{fig:sim1D}, left). However, for logarithmic and sinusoidal functions (Figure \ref{fig:sim1C}, middle and right), extrapolation accuracy diminishes in sparsely supported regions, increasing absolute bias for all estimands (Figure \ref{fig:sim1D}, middle and right).

The FDRC achieves lower absolute bias, particularly when outcome functions are misspecified and extrapolation is unreliable. By focusing on well-supported regions, it avoids large biases and maintains robust estimates even in areas with low non-overlap. Conversely, the trimming estimand exhibits high variability and substantial absolute bias, even in the linear case, due to its restrictive population definition.

These results emphasize the benefits of the most feasible intervention approach, which balances estimation reliability and accuracy by limiting extrapolation only in feasible intervention regions.

\subsubsection{Simulation 2}

\begin{figure*}
    \centering
    \subfloat[True conditional density distribution of $A|L$ in simulation 2. \label{fig:sim2A}]{\includegraphics[width=0.47\textwidth]{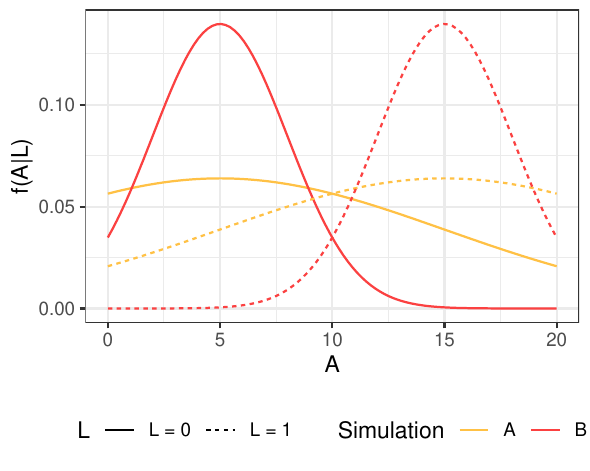}}
    \hspace{0.2cm}
    \subfloat[Non-overlap ratio: the conditional support with support level of $95\%$. \label{fig:sim2B}]{\includegraphics[width=0.47\textwidth]{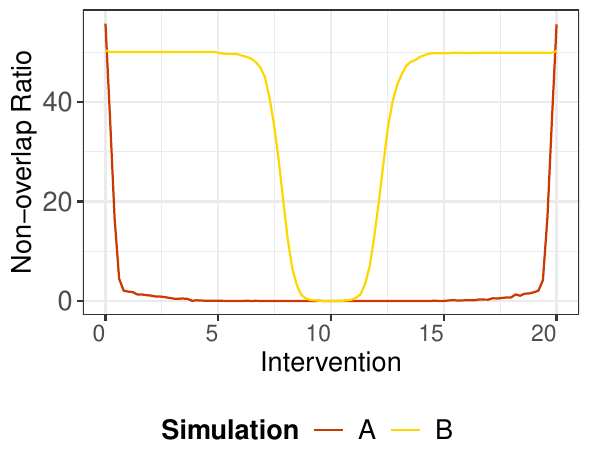}}
    
    \subfloat[Comparison of true and estimated values of \(m^\text{standard}\), \(m^\text{feasible}\), and \(m^\text{trimming}\) from Simulation 2. Separated by Simulation 2A and 2B. \label{fig:sim2C}]{\includegraphics[width=0.95\textwidth]{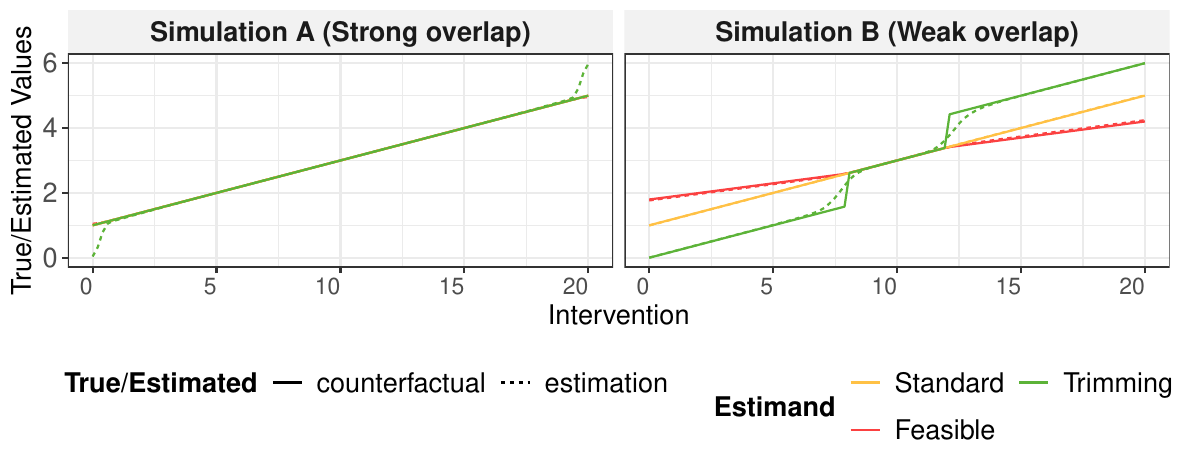}}
    
    \subfloat[Absolute bias of the estimation of \(m^\text{standard}\), \(m^\text{feasible}\), and \(m^\text{trimming}\) from Simulation 2. Separated by Simulation 2A and 2B. \label{fig:sim2D}]{\includegraphics[width=0.95\textwidth]{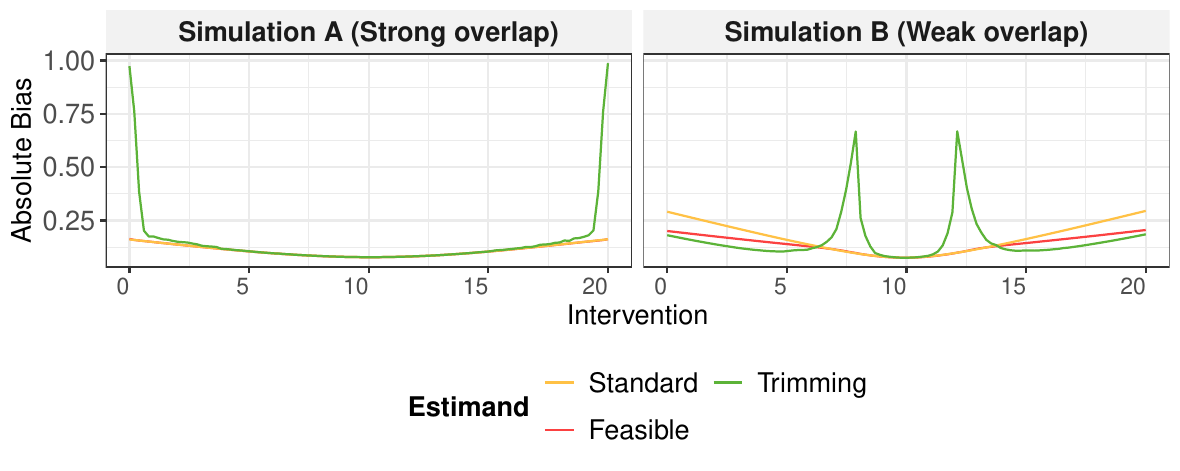}}
    \caption{Results of Simulation 2}
    \label{res:sim2}
\end{figure*}

Figure \ref{fig:sim2A} illustrates the conditional distribution of \( A \) given \( L \). In Simulation 2A, this distribution is more concentrated, with lower variance compared to Simulation 2B. This distribution design increases the chance of non-overlap in Simulation 2B, as shown in Figure \ref{fig:sim2B}.

Figure \ref{fig:sim2B} presents the non-overlap ratio for each intervention. In Simulation 2B, the ratio is lower in the central region but rises sharply to 50\% in the extreme intervention regions due to the divergence between the conditional distributions of \( A \) for \( L = 0 \) and \( L = 1 \). In contrast, Simulation 2A exhibits generally low non-overlap ratios across intervention values, except near the boundaries. This difference arises because, in Simulation 2B, the intervention distributions for \( L = 0 \) and \( L = 1 \) are distinct, whereas in Simulation 2A, they are more similar, exhibiting higher overlap.

Figure \ref{fig:sim2C} shows the true estimand values for both the standard CDRC, FDRC and trimming estimand. In Simulation 2A, where non-overlap is absent, the three true curves are nearly identical because the intervention is feasible for all individuals. However, in Simulation 2B, the estimands diverge outside the central region, as the most feasible intervention shifts toward the center due to limited feasibility at extreme values, and the trimming estimand exhibits more radical counterfactual shifts. For all three curves, the estimates are very close to the true values, as the true outcome function is designed to be straightforward.

Figure \ref{fig:sim2D} compares the absolute bias between Simulations 2A and 2B. Compared to Simulation 2A, Simulation 2B exhibits positivity violation issues, leading to higher absolute bias for all estimands, particularly in regions with high non-overlap. Nonetheless, the FDRC shows lower bias under both conditions compared to the standard estimand and does not exceed the bias level of the standard estimand when positivity violations are minimal, as observed in Simulation 2A. The trimming estimand performs well at times, but it is unstable in certain areas.

\subsubsection{Simulation 3}

\begin{figure*}
    \centering
    \subfloat[Non-overlap ratio: the conditional support with support level of $95\%$. \label{fig:sim3A}]{\includegraphics[width=0.48\textwidth]{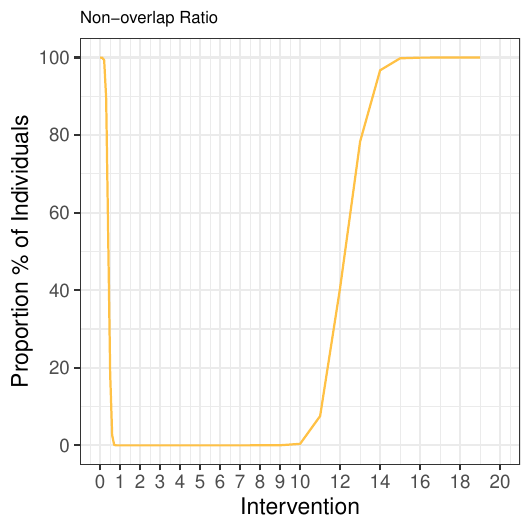}}
    \hspace{0.2cm}
    \subfloat[Comparison of true and estimated values of \(m^\text{standard}\), \(m^\text{feasible}\), and \(m^\text{trimming}\) from Simulation 3. \label{fig:sim3B}]{\includegraphics[width=0.48\textwidth]{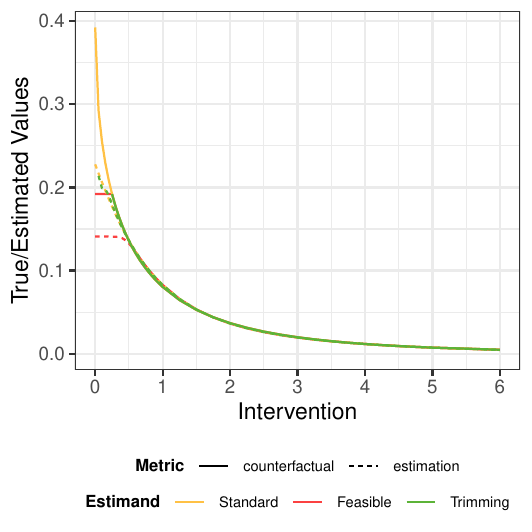}}
    
    \subfloat[Absolute bias of the estimation of \(m^\text{standard}\), \(m^\text{feasible}\), and \(m^\text{trimming}\) from Simulation 3. \label{fig:sim3C}]{\includegraphics[width=0.5\textwidth]{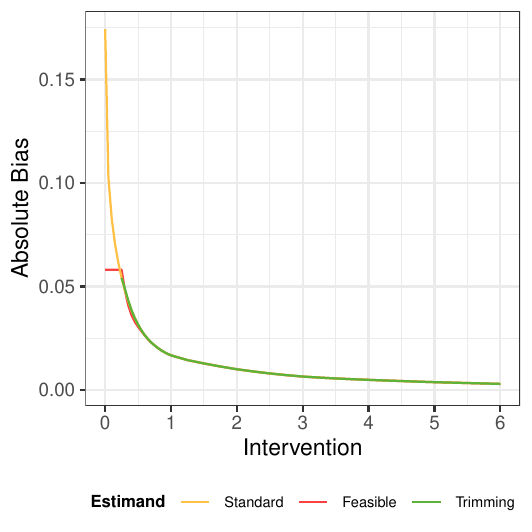}}
    \caption{Results of Simulation 3}
    \label{res:sim3}
\end{figure*}

Figure \ref{fig:sim3A} illustrates the non-overlap ratio in the range of $[0,20]$, which shows a sharp increase at very low intervention values below $0.7$ and above $10$. In the following, we focus on the intervention region of $[0,6]$.

Figure \ref{fig:sim3B} presents the true and estimated values for the three estimands. In regions where the non-overlap ratio is zero, the three estimated curves are indistinguishable. However, the standard CDRC increases dramatically near an intervention value of $0$. As indicated by the non-overlap ratio, such values are nearly impossible to intervene on. In the data-generating process (see Appendix \ref{apd:sim3}), the truncated normal distribution excludes values below $0.2032$, causing the true curve of the trimmed estimand to terminate near this intervention value. This indicates that intervention values below $0.2032$ are strictly infeasible, representing a theoretical positivity violations and thus making counterfactual curve of trimming estimand unidentifiable below that threshold. Unlike the standard CDRC, the proposed FDRC does not exhibit extreme values in regions with very low intervention, making it more robust and less prone to positivity issues.

Figure \ref{fig:sim3C} demonstrates that the FDRC consistently achieves lower absolute bias, particularly in regions with positivity violations. It effectively avoids large biases and maintains robust estimates even in areas with low or no non-overlap. While the trimming estimand also reduces bias, it becomes not estimable when the non-overlap ratio approaches $1$, as no samples are available in such scenarios.

These findings underscore the practical utility of the proposed methods for complex data-generating processes with real-world significance.

\section{Data Analysis}\label{sec:analysis}

We demonstrate our approach using pharmacoepidemiological data of HIV-positive children from the CHAPAS-3 trial, which enrolled children aged 1 month to 13 years in Zambia and Uganda \cite{Mulenga:2016, Abongomera:2017}. We focus on those 125 children who received efavirenz as part of antiretroviral therapy, together with both lamivudine and either stavudine, zidovudine, or abacavir.

In line with previous studies \cite{bienczak2016plasma, bienczak2017determinants, Schomaker:2024b}, we are interested in
how counterfactual viral failure probabilities (VL$_t$), defined as a viral load exceeding $100$ copies/mL, vary as a function of different efavirenz concentrations (EFV$_t$, defined as the plasma concentration (in mg/L) measured $12$ hours after dosing). Briefly, different patients who take the same efavirenz dose, may still have different drug concentrations in their blood, for example because of their individual metabolism. Patients exposed to suboptimal concentrations may experience negative outcomes and one may ask at which concentration level antiretroviral activity is too low to guarantee suppression for most patients. Thus, one may be interested in the concentration-response curve $a \mapsto P(\text{VL}^{\text{EFV}=a}=1), \, a \in [0,6]$ mg/L; although it is known that not every child can achieve every possible concentration level, depending on their clinical and metabolic profile and thus positivity violations may be likely \cite{Schomaker:2024b}. 

For illustration, we use baseline data and data from the scheduled visit at \(t=36\) (weeks). The relevant part of the assumed data-generating process follows previous studies \cite{Schomaker:2024b, Holovchak:2024, Schomaker:2024}.

Our analysis illustrates the ideas based on a complete case analysis of all measured variables at $t=36$ (weeks) represented in the DAG $(n=83)$. The analysis was conducted using three support levels for detecting non-overlap: $99\%$, $95\%$, and $90\%$. Our reference target estimand is the standard causal concentration-response curve, defined in Section \ref{estmand:1}. We compare this estimand to the proposed FDRC, described in Section~\ref{sec:feasible_estimand}, as well as the trimming estimand in Section \ref{estmand:3}. The estimation process closely follows the detailed methodology described in Section \ref{sec:estimation} and algorithm outlined in Algorithm \ref{alg:feasible}. 


\subsection{Results}

\begin{figure}[htbp]
    \centering
    \subfloat[Non-overlap ratio at support level of $90\%$, $95\%$ and $99\%$. As the support level increases, there is greater tolerance for non-overlap detection. Major positivity violations occur in intervention regions with values below $1$ and above $3.5$. \label{fig:ana1}]{\includegraphics[width=0.48\textwidth]{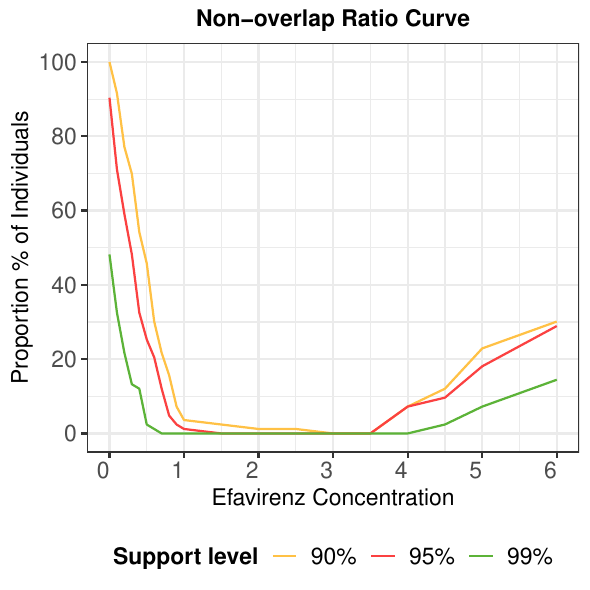}}
    \hspace{0.2cm}
    \subfloat[Comparison of different curves including feasible interventions with support level $90\%$, $95\%$ and $99\%$, and the standard curve. \label{fig:ana2}]{\includegraphics[width=0.48\textwidth]{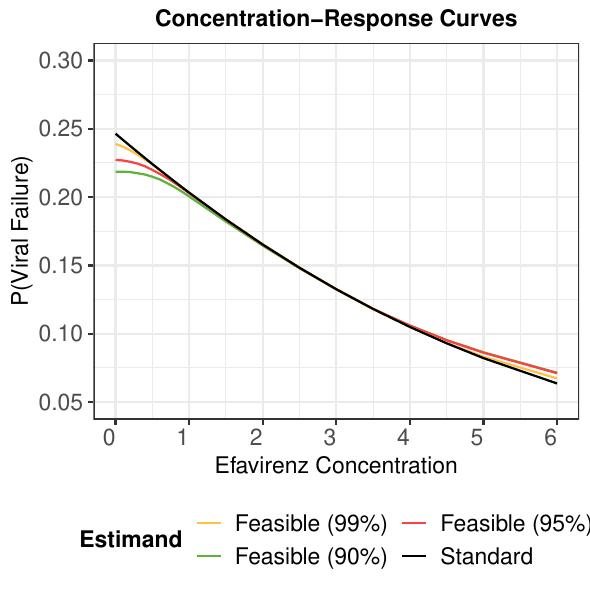}}

    \subfloat[Comparison with of different curves. Separated by support level of $90\%$, $95\%$ and $99\%$. \label{fig:ana3}]{\includegraphics[width=0.95\textwidth]{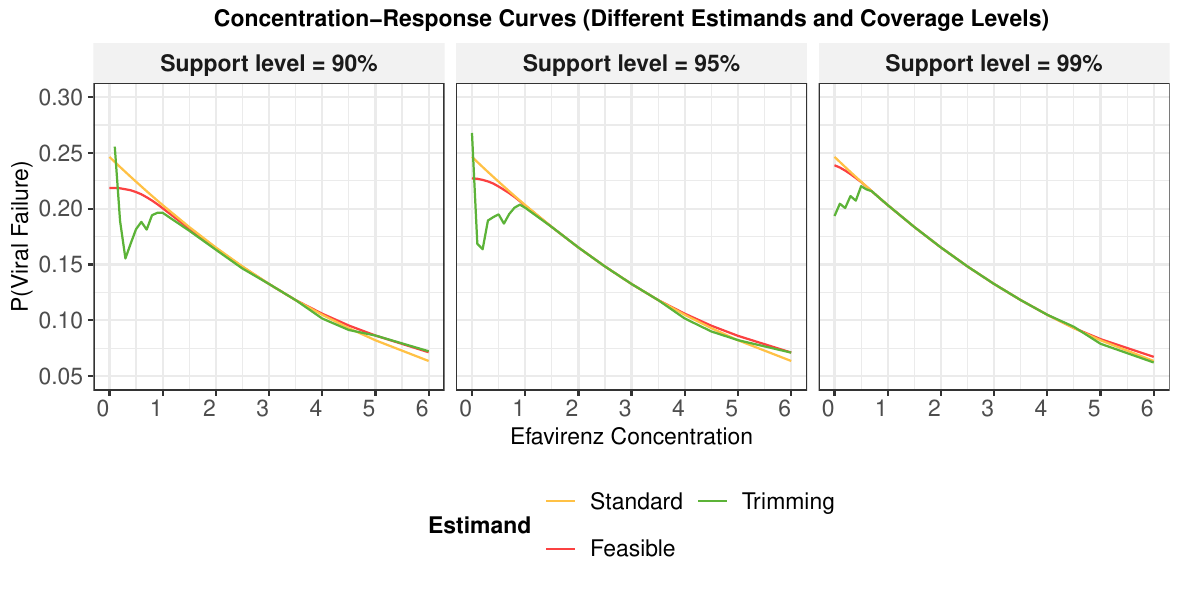}}
    \caption{Results of the data analysis. In panels (b)--(c), the x-axis shows the efavirenz concentration level we \emph{aim for} (the target/attempted concentration). The feasible curve answers: ``If we try to set everyone to this concentration, but for some children it is not realistically attainable, what would the outcome look like after moving those children to their nearest attainable concentration?'' The trimmed curve answers: ``What would the outcome look like at this concentration among the children for whom this concentration appears supported by the data?'' The standard CDRC is shown at the same target concentration values for comparison.}
    \label{ana:res}
\end{figure}

The results of the analysis are presented in Figure \ref{ana:res} and are discussed in two parts: (1) diagnostics assessing the extent of positivity violations based on intervention values, and (2) estimates of three estimands.

Figure \ref{fig:ana1} reports the proposed non-overlap ratio as a diagnostic for positivity violations under continuous interventions. For each target intervention value \(\text{EFV}_{36}=a\), the ratio estimates the fraction of children for whom that target appears unsupported given their baseline covariates, shown for three support levels (\(90\%\), \(95\%\), and \(99\%\)).

Across the central range of concentrations (approximately \(1\) to \(3.5\) mg/L), the non-overlap ratio is close to zero at all support levels, indicating that these targets are broadly supported and that causal estimation is least sensitive to positivity issues. In contrast, the ratio increases at the boundaries—below \(1\) mg/L and above \(3.5\) mg/L—indicating that these targets are unsupported for a nontrivial subset of the population. The sharp rise near \(0\) mg/L is consistent with concentrations in this region being extremely rare in the data.

Figure \ref{fig:ana2} shows the estimated feasible dose--response curves at three support levels alongside the standard causal concentration--response curve (CDRC). In well-supported regions the feasible and standard curves are similar, while they diverge near the boundaries where non-overlap is detected. This behavior reflects the definition of the feasible strategy: for a given target concentration on the x-axis, children for whom that target is unsupported are mapped to the closest supported concentration before evaluating the outcome regression. In this setting, such adjustments are most relevant for extreme targets that some clinical/metabolic profiles are unlikely to attain.

Figure \ref{fig:ana3} compares all estimated curves by support level. As the support level increases, the procedure tolerates more non-overlap, so feasible and trimming curves can differ over a wider range of targets. In this analysis, the estimated trimming curve becomes unstable in regions with strong non-overlap, suggesting limited practical support at the extremes.

Overall, in this HIV application the feasible curve should be read as an \emph{attempted-concentration} curve: the x-axis is the EFV\(_{36}\) concentration level we try to set for everyone. For each attempted value \(a\), the curve answers the following question: if we aim for \(a\) for every child, but \(a\) is not realistically attainable for some children given their covariates (as indicated by the estimated non-overlap), what would the average viral-failure probability look like after assigning those children to their nearest attainable concentration? The resulting curve is plotted against the attempted value \(a\), so differences from the standard CDRC occur primarily where the data indicate that the literal intervention ``set \(A=a\) for everyone'' is not well supported.

\section{Discussion}\label{sec:doscission}
\subsection{Summary}\label{dis:summary}

Positivity violations are a persistent challenge in causal inference with continuous interventions, particularly particularly when estimation requires extrapolation from observed to unobserved regions. Such violations arise when certain intervention levels are insufficiently represented in the data, undermining both the identification and estimation of causal effects.

We address this challenge by using an \(\vb L\)-specific density cutoff \(f_\alpha(\vb l)\) to define conditional highest-density support sets \(\mathcal A_\alpha(\vb l)\). This is closely related in spirit to Modified Treatment Policies (MTPs): both adapt interventions to the observed support structure and therefore reduce reliance on extrapolation in low-overlap strata. The distinction is in the target and construction. MTPs typically define a prespecified shift/transform of treatment (anchored to observed treatment values), while our framework is indexed by a target level a and then enforces feasibility by restricting or projecting to \(\mathcal A_\alpha(\vb l)\), with the goal of preserving comparability to the standard CDRC as much as possible. As emphasized by Ring et al. \cite{ring2025diagnostic}, positivity concerns for MTPs can remain if the induced intervention values themselves are weakly supported; this same principle motivates our explicit non-overlap diagnostics and HDR-based restriction.

Our approach also contrasts with the weighted CDRC proposed by Schomaker et al.\ \cite{Schomaker:2024}. Both approaches address positivity violations via conditional densities, but they differ in how the estimand is adapted. First, we use a covariate-specific support set (indexed by $\alpha$) rather than a fixed cutoff $c$, whereas Schomaker et al.\ use a fixed cutoff \(c\), which may be unsuitable across intervention ranges and covariate strata. Second, our framework serves dual purposes: it provides both overlap diagnostics (see Section~\ref{sec:diagnostic}) and a principled way to define a feasible intervention rule. A further distinction lies in interpretability. Our method modifies intervention assignments explicitly (to the ``closest feasible'' value) and targets the clearly defined quantity \(E(Y^{d_{0,\alpha}(a,\cdot)})\). In contrast, the weighted CDRC redefines the estimand implicitly through a projection inside the expectation, which can make the target parameter less transparent.

Suppose, for example, that \(f(0.5 \mid \text{adherent}, \text{slow metabolizer}) \approx 0\), indicating that assigning treatment \(a=0.5\) is biologically implausible for that stratum. The weighted CDRC of Schomaker et al.\ addresses this by \emph{redefining the target} at \(a\) through a projection that downweights such strata. The identified parameter coincides with the standard causal curve in well-supported strata \(\{f(a\mid \vb{l})>c\}\), while replacing contributions from unsupported strata \(\{f(a\mid \vb{l})\le c\}\) with a weighted associational term. In this example, the \((\text{adherent}, \text{slow metabolizer})\) group contributes negligibly at \(a=0.5\) because \(f(0.5\mid \vb{l})\approx 0\); identification can be maintained under weaker overlap conditions. The trade-off is that \(m^{\text{weighted}}(a)\) no longer corresponds to a clear assignment target such as \(\mathbb{E}(Y^a)\) or \(\mathbb{E}(Y^{d})\): it blends the causal curve where support exists with a projection-based contribution elsewhere, thereby shifting the scientific estimand; the resulting interpretation depends on the weighting/projection in low-support regions. By contrast, our method substitutes infeasible interventions with a feasible rule, retaining a transparent intervention-based interpretation.

A key consequence of limited overlap is increased finite-sample bias and sensitivity to modeling choices, because inference effectively relies on extrapolation into poorly supported regions. Thus, the problem is not only model misspecification but limited empirical support; under strict violations, identification may fail altogether. Our intervention framework addresses this challenge by reassigning infeasible interventions to regions with adequate support, thereby reducing reliance on extrapolation and typically improving finite-sample performance.

\subsection{Limitations}
\label{sec:limitation} 

Estimating the full CDRC under continuous interventions faces several important limitations. First, under mild smoothness conditions, the CDRC is non-pathwise differentiable \cite{diaz2013targeted}. This makes it more challenging to apply standard influence-function-based estimators such as TMLE or doubly robust procedures in the fully nonparametric model, and typically motivates introducing additional structure. Several recent approaches obtain robustness by leveraging assumptions such as differentiability \cite{kennedy2017non}, monotonicity \cite{westling2020causal}, or smooth invertibility \cite{diaz2023nonparametric}. While these assumptions can be reasonable in many applications, they may not always hold or may require careful justification depending on the scientific context. Accordingly, achieving double robustness or semiparametric efficiency often relies on working under a more restricted model.  

Second, our proposed feasible intervention strategy introduces dependence between the observed data and the estimand. As noted by Hubbard et al.\ \cite{hubbard2016statistical}, this dependence falls outside the conditions typically used to justify standard resampling arguments (e.g., the nonparametric bootstrap), and can affect coverage in finite samples. Sample splitting or cross-validation can reduce this dependence by separating the construction of the data-based intervention rule from estimation of the target parameter, but these remedies introduce practical complexity: they can lead to split-specific versions of the estimand and complicate interpretation.

Third, the performance of our approach depends on the stability of estimated intervention rule itself. In particular, the choice of the support level \(\alpha\), the method for estimating conditional densities and the highest density regions. Instability in these components may introduce variability or sensitivity in finite samples, raising concerns about robustness and reproducibility across different applications.

In light of these challenges, our work pursues a complementary goal through the feasible dose–response curve (FDRC): preserve the scientific objective of the original CDRC as closely as possible, while addressing positivity violations via an individualized feasible intervention rule. This yields a natural plug-in estimator that avoids extrapolation and provides valid inference under correct model specification, provided the intervention rule is reasonably stable. Although our approach does not deliver double robustness, it enhances interpretability and it improves interpretability and can reduce sensitivity to modeling choices by focusing estimation on well-supported regions. To evaluate the impact of estimand-data dependency on inference, we conducted a simulation study (Appendix~\ref{apd:bst}) comparing bootstrap coverage with and without sample splitting. The results suggest that, for the scenarios considered and in well-supported regions, bootstrap intervals can be reasonably accurate even without splitting. 

Finally, our method provides a principled approach for substituting infeasible interventions with their most feasible counterparts—thereby retaining an intervention-based interpretation closely related to the standard CDRC. Further research is needed to extend the framework to longitudinal settings and develop more robust techniques for estimating support regions, especially in high-dimensional and complex settings.

\section{Software Implementation in the \texttt{CICI} Package}
\label{sec:software}

\subsection{\texttt{feasible()}: Feasible-Intervention Estimation}

The proposed feasible-intervention method is implemented in the \texttt{R}-CRAN package \texttt{CICI} \citep{Schomaker:2024,manualcici}.  

We illustrate usage with the package dataset \texttt{EFV}, which contains longitudinal treatment, outcome, and covariate measurements. For a one-time-point demonstration, we retain only baseline variables up to \texttt{VL.0}, yielding a one-time-point setup with treatment \texttt{efv.0}, outcome \texttt{VL.0}, and baseline covariates \texttt{sex}, \texttt{metabolic}, \texttt{log\_age}, \texttt{NRTI}, and \texttt{weight.0}. In this reduced setting, there are no longitudinal covariate nodes (\texttt{Lnodes = NULL}).

We then estimate feasible interventions over target values \texttt{abar\_1tp = 0,1,\ldots,10}, summarize feasibility diagnostics, and visualize both the mean feasible mapping and non-overlap region.

\begin{codeblock}
# A.X | One-time-point feasible intervention example (baseline only)

# Step 0: Load package and example data
# install.packages("CICI")
library(CICI)
data(EFV)

# Step 1: Prepare one-time-point data (baseline through VL.0)
# Variables kept: sex, metabolic, log_age, NRTI, weight.0, efv.0, VL.0
EFV_1tp <- EFV[, c("sex","metabolic","log_age","NRTI",
                   "weight.0","efv.0","VL.0")]

# Step 2: Define nodes for the one-time-point setting
Anodes <- "efv.0"      # treatment at t = 0
Ynodes <- "VL.0"       # outcome at t = 0 (illustrative baseline outcome)
Lnodes <- NULL         # no longitudinal covariates in this reduced dataset

# Step 3: Define target intervention values
abar_1tp <- seq(0, 10, by = 1)

# Step 4: Estimate feasible interventions
m_1tp <- feasible(
  X = EFV_1tp,
  Anodes = Anodes,
  Ynodes = Ynodes,
  Lnodes = Lnodes,
  abar = abar_1tp,          # target treatment values to evaluate
  alpha = 0.95,             # feasibility level: retain top 95% density mass;
                            # the lowest ~5% tail is flagged as low-overlap
  d.method = "parametric", # conditional density estimator g(a | L)
  grid.size = 0.5,          # internal grid spacing for density approximation
                            # smaller = finer resolution (more computation)
  left.boundary = 0,        # lower boundary of treatment support grid
  right.boundary = 10       # upper boundary of treatment support grid
)

# Step 5: Inspect feasible values and overlap diagnostics
# Each m_1tp$feasible[[k]] is an (n x 1) matrix:
#   rows    = individuals
#   column  = time point (here: efv.0 only)
#   entries = individualized feasible value under strategy k (abar_1tp[k])

head(do.call(cbind, m_1tp$feasible))
    0   1 2 3 4 5 6 7 8 9   10
1 0.0 1.0 2 3 4 5 6 7 8 9  9.5
2 0.0 1.0 2 3 4 5 6 7 8 9  9.0
3 1.5 1.5 2 3 4 5 6 7 8 9 10.0
4 0.0 1.0 2 3 4 5 6 7 8 9  9.0
5 0.0 1.0 2 3 4 5 6 7 8 9  9.0
6 0.0 1.0 2 3 4 5 6 7 8 9  9.5

# Interpretation:
# After cbind, columns correspond to target strategies (abar_1tp = 0,...,10).
# Each cell is the individualized feasible value under that target.
# Values != the target (e.g., target 10 mapped to 9/9.5) indicate limited overlap.

# Strategy-level summaries
m_1tp             # see ?print.feasible
summary(m_1tp)    # see ?summary.feasible

# Step 6: Plot diagnostics
#
# Plot (a): which = "feasible"
#   - x-axis: target intervention value a in abar_1tp
#   - y-axis: mean feasible value E[d(a, .)] across individuals
#
# Plot (b): which = "nonoverlap"
#   - low-density regions under fitted conditional treatment density g(a | L)

plot(m_1tp, which = "feasible")    # mean feasible value across individuals, by target strategy
plot(m_1tp, which = "nonoverlap")  # low-density/non-overlap region

# Step 7: Estimate and plot the feasible dose-response curve (FDRC)
#
# gformula() expects abar to represent the intervention values to plug in.
# Here we pass the individualized feasible interventions returned by feasible().

est_fdrc <- gformula(X = EFV_1tp,
  Anodes = Anodes,
  Ynodes = Ynodes,
  Lnodes = NULL, 
  abar = m_1tp$feasible # use results from Step 4)

plot(est_fdrc)  # see ?plot.gformula

\end{codeblock}

\begin{figure}[htbp]
  \centering

  \subfloat[Mean feasible intervention value (averaged over individuals) by target strategy.\label{fig:software-feasible}]{
    \includegraphics[width=0.48\textwidth]{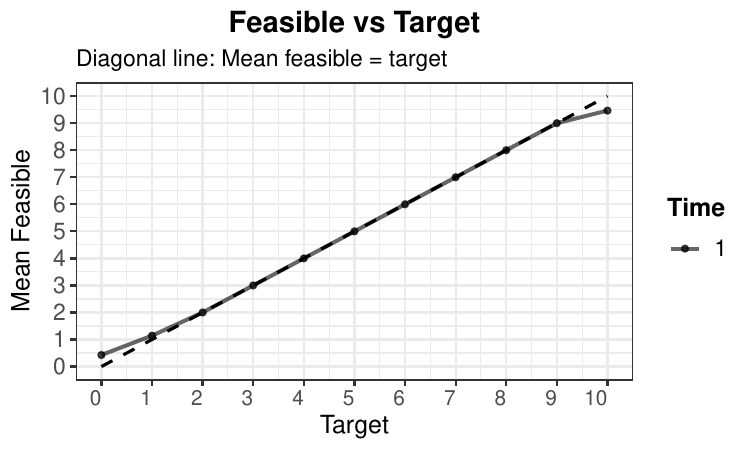}
  }
  \hfill
  \subfloat[Estimated non-overlap (low-density) region under the fitted treatment density.\label{fig:software-nonoverlap}]{
    \includegraphics[width=0.48\textwidth]{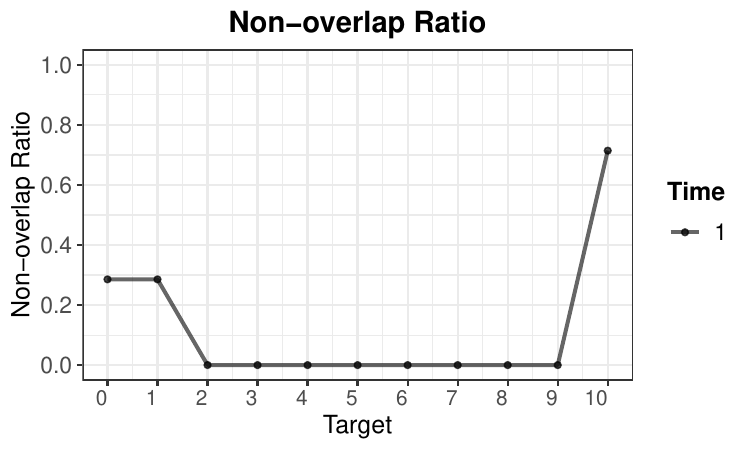}
  }

  \vspace{0.35cm}

  \subfloat[Feasible dose--response curve (FDRC).\label{fig:software-fdrc}]{
    \includegraphics[width=0.70\textwidth]{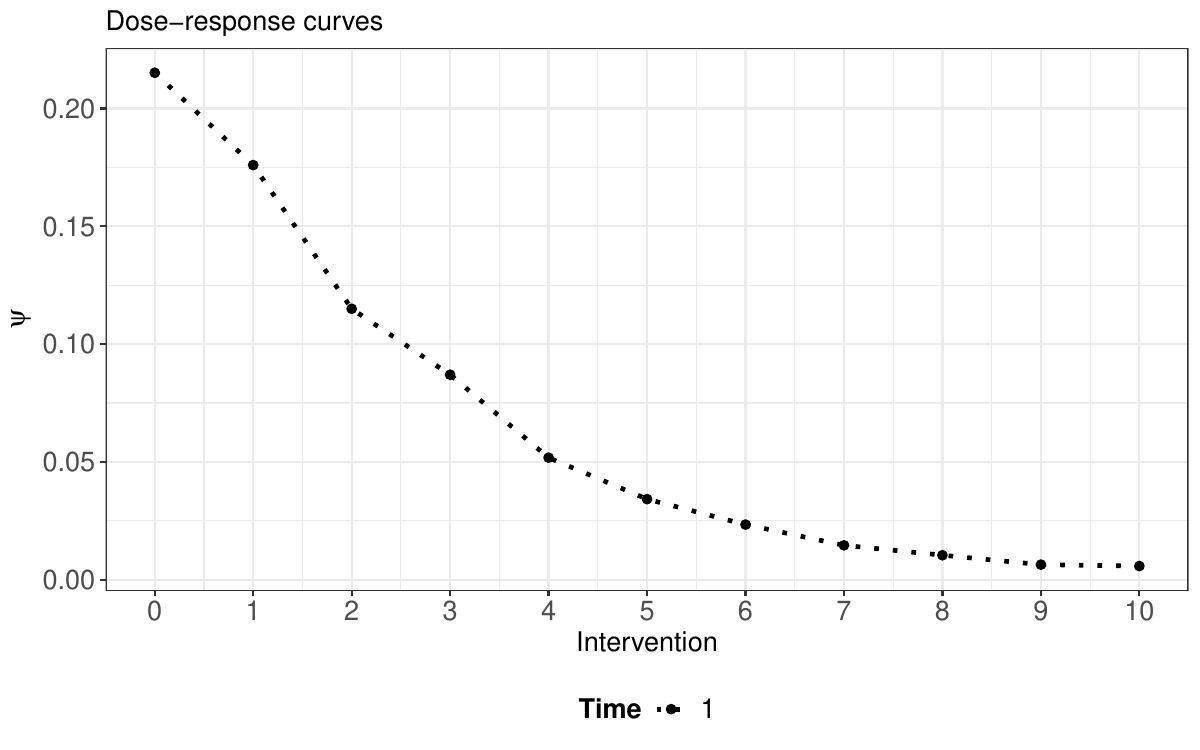}
  }

  \caption{Software illustration for the one-time-point example. Panels (a)--(b) show feasibility diagnostics from \texttt{feasible()}: (a) the mean feasible mapping (average individualized feasible value at each target), and (b) intervention regions that are frequently unrealistic for some individuals (non-overlap ratio). Panel (c) shows the resulting feasible dose--response curve estimated via \texttt{gformula()} using the individualized feasible interventions; the x-axis denotes the target (attempted) intervention level, and outcomes are evaluated after mapping individuals who cannot realistically attain the target to their nearest attainable value.}
  \label{fig:software-panels}
\end{figure}

Figure~\ref{fig:software-feasible} shows the mean feasible mapping, i.e., the average
of individualized feasible values at each target intervention level.
When the curve is close to the 45-degree identity line ($y=x$), target values are largely
supported by the observed data. Systematic deviations from this identity indicate that
targets are being remapped to nearby higher-density values, reflecting practical
positivity constraints. Larger deviations imply stronger constraints.

Figure~\ref{fig:software-nonoverlap} shows treatment regions classified as low-density
(non-overlap) under the fitted $g(a\mid H)$. These are areas with weak empirical support
for direct intervention. Broader flagged regions indicate more severe overlap limitations,
and flagged extremes typically correspond to where the mean feasible mapping departs most
from $y=x$.

Figure~\ref{fig:software-fdrc} shows the estimated feasible dose--response curve (FDRC).
The x-axis represents the \emph{target} intervention value that we aim to set.
At each target value, the FDRC reports the mean counterfactual outcome after assigning
each individual to that target when it is realistically attainable for them, and otherwise
to the closest attainable value returned by \texttt{feasible()}. Consequently, the curve
typically agrees with the standard plug-in curve in well-attainable regions and departs
from it primarily at extreme targets where many individuals are remapped to nearby values.

\subsection{Estimate the Conditional Density via Hazard-Binning}
\label{apd:hazard}
\index{density estimation!HAL}\index{HAL}\index{highly-adaptive LASSO}

Consider discretizing the intervention $a$ of interest into \( K \) bins, which may not necessarily be of the same width. Additionally, we can make the following observation: the probability of $a$ falling into a specific bin equals the probability of falling into the respective bin given that $a$ has not been placed into any of the previous bins times the probability of not falling into any of those those previous bins:

\begin{align}\label{eqn:hazardbinning}
P(a \in [\alpha_{k-1}, \alpha_k) |\mathbf{L}) &= P(a \in [\alpha_{k-1}, \alpha_k) | A \geq \alpha_{k-1}, \mathbf{L}) \times  \nonumber\\
&\ \ \prod_{j=1}^{k-1} \left\{ 1 - P(a \in [\alpha_{j-1}, \alpha_j) | A \geq \alpha_{j-1}, \mathbf{L}) \right\}
\end{align}

Under the assumption that within each interval $[\alpha_k - \alpha_{k-1})$ the density is constant, we can estimate it as follows:

\begin{equation}\label{eqn:hazardbinning_density}
f(a|\mathbf{l}) = \frac{P(a \in [\alpha_{k-1}, \alpha_k) |\mathbf{L})}{\alpha_k - \alpha_{k-1}}, \ \forall a \in [\alpha_k - \alpha_{k-1})
\end{equation}

To obtain the density, we can apply either parametric or non-parametric estimation to the hazard model and then transform the results into a density for each bin. This approach is adapted from Haldensify, proposed by Nima S. Hejazi et al. \cite{hejazi2022haldensify, hejazi2022haldensify-rpkg, hejazi2022haldensify-joss}, and it is similar to the approach described in an earlier proposal \cite{diaz2011super}.

One could estimate the probability of \( a \) falling into the interval \( [\alpha_{k-1}, \alpha_k),\ k=1,\cdots,K \), i.e. the left part of the above equation, using any regression or machine learning model. However, if density values are low in a specific interval this becomes very unstable. A safer option is to estimate the right hand side of (\ref{eqn:hazardbinning}) because the products of ``\textit{one minus} the conditional probabilities'' impliy multiplication with ``1'' in low density intervals. To practically estimate the right hand side of (\ref{eqn:hazardbinning}), one has to re-format the data  and fit a model ``pooled over time'' on it:

{\small
\begin{verbatim}
## Example of "pooled over bins" data for hazard-binning
## (three subjects, K = 3 bins on A)

id   A    L1    L2   k   alpha_start   alpha_end   in_bin
1  0.30  0.20  1.00  1      0.00         0.25        0
1  0.30  0.20  1.00  2      0.25         0.50        1
2  0.62 -0.50  0.30  1      0.00         0.25        0
2  0.62 -0.50  0.30  2      0.25         0.50        0
2  0.62 -0.50  0.30  3      0.50         0.75        1
3  0.10  1.10 -0.70  1      0.00         0.25        1
\end{verbatim}
}

This data setup estimates the right-hand side of (\ref{eqn:hazardbinning}), specifically \(P(a \in [\alpha_{k-1}, \alpha_k) \mid A \geq \alpha_{k-1}, \mathbf{L})\) due to the implicit conditioning set:  at each bin $k$, the pooled data restrict the sample to individuals with
$A>\alpha_{k-1}$; thus, the regression model is fit on the corresponding risk set, and the estimated probabilities are conditional on ``survival'' through all previous bins. This mirrors discrete-time survival analyses, with $a$ playing the role of an event time and \texttt{in\_bin} serving as the bin-specific failure indicator.

Note that with this reshaping the number of rows increases from \(n\) to \(\sum_{i=1}^n K_i\), where \(K_i\) is the number of bins for which subject \(i\) is still at risk (i.e.\ all \(k\) with \(\alpha_{k-1} \le A_i\)); in the worst case this is \(nK\). In the implementation, we therefore report this pooled row count as an ``effective sample size'' when \texttt{feasible()} is called with \texttt{verbose = TRUE} and \texttt{method = "hazardbinning"}.

The \texttt{"hazardbinning"} estimate the right-hand side of (\ref{eqn:hazardbinning}) through fitting pooled models on the reshaped data, then calculate the left-hand side of (\ref{eqn:hazardbinning}) and convert to density estimates (\ref{eqn:hazardbinning_density}).

\section*{Acknowledgments}
We sincerely thank the CHAPAS-3 trial team for their support, valuable advice on the illustrative data analysis, and for providing access to their data.

We extend our special thanks to Sarah Walker, Di Gibb, Andrzej Bienczak, David Burger, Elizabeth Kaudha, Paolo Denti, and Helen McIlleron for valuable advice regarding the data and continuous support. Additionally, we are deeply grateful to Iván Díaz for his constructive and insightful feedback on earlier versions of this manuscript.

The research is supported by the German Research Foundations (DFG) Heisenberg Program (grants 465412241 and 465412441).

\bibliographystyle{plainnat} 
\bibliography{main}   

\appendix
\section{Additional Information on Positivity Violation}\label{apd:pos}

In the literature, positivity violations are typically classified into structural and practical forms, usually for binary interventions \cite{zivich2022positivity, petersen2012diagnosing, westreich2010invited}. The same ideas extend to continuous interventions.

\paragraph{Near positivity violations.}
Near violations occur when positivity is not exactly violated but is close to failing: some treatment levels have conditional probability (or density) very close to zero for certain confounder patterns. Although not a strict violation, the practical consequences are similar---loss of precision and potential instability in causal estimation, especially in sparse regions \cite{leger2022causal}.

\paragraph{Structural positivity violations.}
Structural violations arise when some treatment values are impossible for certain confounder strata by design or mechanism. In continuous-treatment notation, this corresponds to
\[
f_{A\mid \vb L}(a\mid \vb L)=0
\]
for those strata and treatment values.

\paragraph{Practical positivity violations.}
Practical (stochastic) violations occur when treatment values are clinically possible but poorly represented in the observed sample. In finite samples this is reflected through estimated support, e.g. \(\hat f_{A\mid \vb L}(a\mid \vb L)\) near zero. For continuous treatments, this depends on density-estimation choices because exact treatment values are not repeatedly observed.

\bigskip
\noindent\textbf{Theoretical implications for plug-in estimation.}

Positivity violations---whether structural (deterministic lack of support) or practical (data sparsity)---create core challenges for plug-in estimation of the causal dose--response curve (CDRC) in Equation~\eqref{eqn:estimand1}. Define
\[
Q_0(a,\vb l):=E_{P_0}(Y\mid A=a,\vb L=\vb l).
\]
Under standard identification assumptions, the CDRC is
\begin{equation}\label{eqn:cdrc}
m(a)=E_{P_0}\!\left[Q_0(a,\vb L)\right].
\end{equation}

\noindent\textit{Step 1: Outcome-regression fitting.}

In practice, \(Q_0\) is estimated by fitting \(\hat Q(A,\vb L)\) (e.g., least squares). For a candidate \(Q\), the population risk is
\begin{equation}\label{eqn:conditional}
R(Q)=E_{P_0}\!\left[(Y-Q(A,\vb L))^2\right].
\end{equation}
Using variance decomposition:
\[
R(Q)
=
E_{P_0}\!\left[(Q_0(A,\vb L)-Q(A,\vb L))^2\right]
+
E_{P_0}\!\left[\Var(Y\mid A,\vb L)\right].
\]
Hence minimizing \eqref{eqn:conditional} is equivalent to minimizing
\[
\int \big(Q_0(a,\vb l)-Q(a,\vb l)\big)^2\,dP_{A,\vb L}(a,\vb l).
\]
So squared errors are weighted by
\[
f_{A,\vb L}(a,\vb l)=f_{A\mid \vb L}(a\mid \vb l)\,f_{\vb L}(\vb l).
\]
Therefore, low-support regions (small \(f_{A\mid \vb L}(a\mid \vb l)\)) have little influence on the fitted regression.

\noindent\textit{Step 2: Plug-in estimation of the CDRC.}

The plug-in estimator is
\[
\hat m(a)=E_{P_n}\!\left[\hat Q(a,\vb L)\right].
\]
Its mean-squared error at fixed \(a\) is
\begin{equation}\label{eqn:mse}
\mathrm{MSE}\!\big(\hat m(a)\big)
=
E_{P_0}\!\left[\big(\hat Q(a,\vb L)-Q_0(a,\vb L)\big)^2\right]
=
\int \big(\hat Q(a,\vb l)-Q_0(a,\vb l)\big)^2\,dP_{\vb L}(\vb l).
\end{equation}

\noindent\textit{Key mismatch.}

Equations~\eqref{eqn:conditional} and \eqref{eqn:mse} imply:
\begin{itemize}
\item Regression fitting minimizes error under \(P_{A,\vb L}\), i.e., weighted by \(f_{A\mid \vb L}(a\mid \vb l)\,f_{\vb L}(\vb l)\).
\item The CDRC target at fixed \(a\) averages over \(P_{\vb L}\), i.e., weighted only by \(f_{\vb L}(\vb l)\), regardless of local support at \(a\).
\end{itemize}

Equivalently,
\[
R(Q)=\int \big(Q_0(a,\vb l)-Q(a,\vb l)\big)^2 f_{A\mid \vb L}(a\mid \vb l)f_{\vb L}(\vb l)\,da\,d\vb l,
\]
whereas
\[
\mathrm{MSE}\!\big(\hat m(a)\big)
=\int \big(Q_0(a,\vb l)-\hat Q(a,\vb l)\big)^2 f_{\vb L}(\vb l)\,d\vb l.
\]
Thus, when \(f_{A\mid \vb L}(a\mid \vb l)\approx 0\), errors in \(Q_0(a,\vb l)\) receive little penalty during regression fitting but can still contribute substantially to plug-in bias in \eqref{eqn:mse}.

\noindent\textit{Implication.}
This mismatch explains why plug-in g-computation can be sensitive to positivity violations: extrapolation errors in sparse or unsupported strata may be weakly controlled by regression risk yet still strongly affect estimation of \(m(a)\).

\section{Simulation Parameters}

\subsection{Data Generating Process for Simulation 1}\label{apd:sim1}

\begin{equation*}
\label{sim:sim1}
\begin{aligned}
    L &\sim N \left( \mu = 3, \sigma = 1 \right) \\ 
    A &\sim N \left( \mu = 10+0.7 \cdot L, \sigma = 3 \right) 
\end{aligned}
\end{equation*}

\bmsubsubsection{Simulation 1A: Linear outcome function}
\begin{align*} 
    Y &\sim N \left( \mu = 10+A + 0.5\cdot L, \sigma = 3 \right)
\end{align*}

\bmsubsubsection{Simulation 1B: Sinusoidal outcome function}
\begin{align*} 
    Y &\sim N \left( \mu = 10+10\cdot\sin\left(\dfrac{A-10}{10}\right) + 0.5\cdot L, \sigma = 3 \right)
\end{align*}

\bmsubsubsection{Simulation 1C: Logarithmic outcome function}
\begin{align*} 
    Y &\sim N \left( \mu = 10+30\cdot\log\left(\dfrac{A-10}{30}\right) + 0.5\cdot L, \sigma = 3 \right)
\end{align*}

\subsection{Data Generating Process for Simulation 2}\label{apd:sim2}

\begin{equation*}
\label{sim:sim2}
    L \sim \mathrm{Bernoulli} \left( p = 0.5 \right)
\end{equation*}

\bmsubsubsection{Simulation 2A}
\begin{equation*}
\begin{aligned}
    A | L \sim \mathcal{N}_\text{trunc}(&\mu = 10 + 5\cdot L, \sigma = 3, l = 0, h = 20,\\
    &l_a=h_a=0, l_b=h_b=20) \\
    Y |A, L \sim N(&\mu = 0.05\cdot A + 0.5\cdot L, \sigma = 2)
\end{aligned}
\end{equation*}

\bmsubsubsection{Simulation 2B}

\begin{equation*} 
\begin{aligned}
    A | L \sim \mathcal{N}_\text{trunc} (&\mu = 10 + 5\cdot L, \sigma = 10,l = 0, h = 20,\\
    &l_a=h_a=0, l_b=h_b=20) \\
    Y |A, L \sim N(&\mu = 0.05\cdot A + 0.5\cdot L, \sigma = 2)
\end{aligned}
\end{equation*}

\subsection{Data Generating Process for Simulation 3}\label{apd:sim3}

\begin{align*}
\text{Sex} & \sim \text{Bernoulli}(p = 0.5) \\
\text{Genotype} & \sim \text{Categorical} \\ 
&\left(p =
\begin{cases} 
\text{logit}^{-1}(-0.103 + \text{Sex} \cdot 0.223 + \\ 
(1-\text{Sex}) \cdot 0.173), \\ 
\text{logit}^{-1}(-0.086 + \text{Sex} \cdot 0.198 + \\ 
(1-\text{Sex}) \cdot 0.214), \\ 
\text{logit}^{-1}(-0.309 + \text{Sex} \cdot 0.082 + \\ 
(1-\text{Sex}) \cdot 0.107)
\end{cases}\right) \\
\text{Age} & \sim \mathcal{N}_\text{trunc}\big(\mu = 1.501, \sigma = 0.369, l = 0.693, h = 2.8, \\
&l_a = 0.693, l_b = 1, h_a = 1, h_b = 2.8) \\
\text{Weight}_{0} & \sim \mathcal{N}_\text{trunc}\big( 
\mu = 1.5 + 0.2 \cdot \text{Sex} + 0.774 \cdot \text{Age}, \\
& \quad \sigma = 0.3, l = 2.26, h = 3.37, \\
&l_a = 2.26, l_b = 2.67, h_a = 3.02, h_b = 3.37) \\
\text{NRTI} & \sim \text{Categorical} \\
&\left(p = 
\begin{cases} 
\text{logit}^{-1}(-0.006 + \text{Age} \cdot (0.1735 \cdot \mathbbm{1}_{\{\text{Age} > 1.4563\}} + \\ 0.157)) \\ 
\text{logit}^{-1}(-0.006 + \text{Age} \cdot (0.1570 \cdot \mathbbm{1}_{\{\text{Age} > 1.4563\}} + \\ 0.1818))
\end{cases}
\right) \\
\text{CoMo}_{0} & \sim \text{Bernoulli}(p = 0.15) \\
\text{Dose}_{0} & \sim \text{Categorical} \\
&\left(p = 
\begin{cases} 
\text{logit}^{-1}(5 + 8 \cdot \sqrt{\text{Weight}_{0}} - 10 \cdot \text{Age}) \\ 
\text{logit}^{-1}(4 + 8.768 \cdot \sqrt{\text{Weight}_{0}} - 9.060 \cdot \text{Age}) \\ 
\text{logit}^{-1}(3 + 6.562 \cdot \sqrt{\text{Weight}_{0}} - 8.325 \cdot \text{Age})
\end{cases}
\right) \\
\text{EFV}_{0} & \sim \mathcal{N}_\text{trunc}\big( 
\mu = -8 + 0.1 \cdot \text{Age} + 4.66 \cdot \text{Genotype} \\
& \quad + 0.1 \cdot \text{Dose}_{0} + 2.66 \cdot \mathbbm{1}_{\{\text{Genotype} \leq 2\}} \\
& \quad + 4.6 \cdot \mathbbm{1}_{\{\text{Genotype} = 3\}}, \sigma = 4.06, l = 0.2032, h = 21, \\
&l_a = 0.2032, l_b = 0.88, h_a = 8.376, h_b = 21)\big) \\
\text{VL}_{0} & \sim \text{Bernoulli}\big(p = 1 - \text{logit}^{-1}(0.4 \\
& \quad + 1.9 \cdot \sqrt{\text{EFV}_{0}}) \big) \\
\text{MEMS} & \sim \text{Bernoulli}\big(p = \text{logit}^{-1}(0.31 \cdot \text{CoMo}_{0} \\
& \quad + 0.71) \big) \\
\text{Weight} & \sim \mathcal{N}_\text{trunc}\big( 
\mu = -0.05 \cdot \mathbbm{1}_{\{\text{CoMo}_{0} = 1\}} + 1.04 \cdot \text{Weight}_{0}, \\
& \quad \sigma = 0.4, l = 2.26, h = 3.37, \\
&l_a = 2.26, l_b = 2.473, h_a = 3.2, h_b = 3.37) \big) \\
\end{align*}

\begin{align*}
\text{CoMo} & \sim \text{Bernoulli}\big(p = 1 - \text{logit}^{-1}(0.5 \cdot \mathbbm{1}_{\{\text{CoMo}_{0} = 1\}} \\
& \quad + 0.1 \cdot \text{Age} + 0.1 \cdot \text{Weight}_{0}) \big) \\
\text{Dose} & \sim \text{Categorical} \\
&\left(p = 
\begin{cases} 
\text{logit}^{-1}(4 + 0.5 \cdot \text{Dose}_{0} + \\
4 \cdot \sqrt{\text{Weight}} - 10 \cdot \text{Age}) \\ 
\text{logit}^{-1}(-8 + 0.5 \cdot \text{Dose}_{0} + \\ 
8.568 \cdot \sqrt{\text{Weight}} - 9.060 \cdot \text{Age}) \\ 
\text{logit}^{-1}(20 + 0.5 \cdot \text{Dose}_{0} + \\
6.562 \cdot \sqrt{\text{Weight}} - 18.325 \cdot \text{Age})
\end{cases}
\right) \\
\text{EFV} & \sim \mathcal{N}_\text{trunc}\big( 
\mu = 0.1 \cdot \text{Dose} + 0.1 \cdot \text{MEMS} + 2.66 \cdot \mathbbm{1}_{\{\text{Genotype} \leq 2\}} \\
& \quad + 4.6 \cdot \mathbbm{1}_{\{\text{Genotype} = 3\}}, \sigma = 4.06, l = 0.2032, h = 21.84, \\
&l_a =0.2032, l_b = 0.88, h_a = 8.37, h_b = 21.84) \big) \\
\text{VL} & \sim \text{Bernoulli}\big(p = 1 - \text{logit}^{-1}(0.4 + 0.1 \cdot \text{CoMo} \\
& \quad + 2 \cdot \sqrt{\text{EFV}}) \big)
\end{align*}

The distribution $N_{\text{trunc}}(\mu, \sigma, l, h, l_a, l_b, h_a, h_b)$ represents a truncated normal distribution, where $l$ and $h$ are the truncation thresholds. Values less than $l$ are replaced with random draws from a $U(l_a, l_b)$ distribution, while values greater than $h$ are replaced with random draws from a $U(h_a, h_b)$ distribution. Here, $U$ denotes a continuous uniform distribution.
\section{Computing the HDR Cutoff via a Quantile of the Density-Value Distribution}
\label{app:fa_quantile}

\subsection{Highest density cutoff and Highest density region}
\label{app:fa_quantile:def}

Fix \(\alpha\in(0,1)\) and \(\vb l\). Let \(f_0(a\mid \vb l)\) denote the conditional density of
\(A\mid \vb L=\vb l\) on a compact support \(\mathcal A\subset\mathbb R\). Define the induced
\emph{density-value random variable} \cite{hyndman1996computing}
\[
Z_{\vb l}=f_0(A\mid \vb l),\qquad A\sim P_{A\mid \vb L=\vb l}.
\]
For any threshold \(t\in\mathbb R\), the event \(\{Z_{\vb l}\ge t\}\) is equivalent to
\(\{f_0(A\mid \vb l)\ge t\}\), and therefore
\[
P\!\big(Z_{\vb l}\ge t\mid \vb L=\vb l\big)
=
P\!\big(f_0(A\mid \vb l)\ge t\mid \vb L=\vb l\big)
=
P\!\big(A\in\{a\in\mathcal A:\ f_0(a\mid \vb l)\ge t\}\mid \vb L=\vb l\big).
\]
Thus, selecting \(t\) as a quantile of \(Z_{\vb l}\) yields a density threshold whose associated upper level set
has a prescribed conditional probability content. We define the cutoff as the \((1-\alpha)\)-quantile of \(Z_{\vb l}\):
\begin{equation}
f_\alpha(\vb l)
=
\inf\Big\{t\in\mathbb R:\ P\!\big(Z_{\vb l}\le t\mid \vb L=\vb l\big)\ge 1-\alpha\Big\}.
\label{eq:fa_quantile_pop}
\end{equation}
The highest density region (HDR) at level \(\alpha\) is then given by the corresponding upper level set
\begin{equation}
\mathcal A_\alpha(\vb l;P_0)
=
\big\{a\in\mathcal A:\ f_0(a\mid \vb l)\ge f_\alpha(\vb l)\big\}.
\label{eq:hdr_closed_def}
\end{equation}
By construction,
\[
P\!\left(A\in\mathcal A_\alpha(\vb l;P_0)\mid \vb L=\vb l\right)
=
P\!\left(Z_{\vb l}\ge f_\alpha(\vb l)\mid \vb L=\vb l\right)
\;\ge\;\alpha,
\]
with equality whenever there is no probability mass at the cutoff, i.e.,
\(P\!\left(Z_{\vb l}=f_\alpha(\vb l)\mid \vb L=\vb l\right)=0\). In the presence of ties
(\(P(Z_{\vb l}=f_\alpha(\vb l)\mid \vb L=\vb l)>0\)), the closed-set convention in
\eqref{eq:hdr_closed_def} yields probability content at least \(\alpha\), which is the natural analogue of
quantile ties in the continuous/discrete boundary case.

\subsection{Grid-based sample analogue}
\label{app:fa_quantile:sample}

Let \(\hat f_n(a\mid \vb l)\) be an estimated conditional density.
Let \(\overline{\vb a}^{\,\mathrm{hdr}}=(a^{\mathrm{hdr}}_1,\dots,a^{\mathrm{hdr}}_{m_{\mathrm{hdr}}})^\top\) be a
strictly increasing grid on \(\mathcal A\), i.e.,
\(a^{\mathrm{hdr}}_1<\cdots<a^{\mathrm{hdr}}_{m_{\mathrm{hdr}}}\).
Define grid cell boundaries \((b_k)_{k=1}^{m_{\mathrm{hdr}}+1}\) by
\[
b_{1}:=a^{\mathrm{hdr}}_{1},\qquad
b_{k}:=\frac{a^{\mathrm{hdr}}_{k-1}+a^{\mathrm{hdr}}_{k}}{2}\ \ (k=2,\dots,m_{\mathrm{hdr}}),\qquad
b_{m_{\mathrm{hdr}}+1}:=a^{\mathrm{hdr}}_{m_{\mathrm{hdr}}}.
\]
Let the associated partition of \([a^{\mathrm{hdr}}_1,a^{\mathrm{hdr}}_{m_{\mathrm{hdr}}}]\) be given by
\[
I_k := [b_k,b_{k+1}) \quad (k=1,\dots,m_{\mathrm{hdr}}-1), 
\qquad
I_{m_{\mathrm{hdr}}}:=[b_{m_{\mathrm{hdr}}},b_{m_{\mathrm{hdr}}+1}],
\]
and define the cell widths as Lebesgue lengths
\[
w_k := \lambda(I_k)=b_{k+1}-b_k,\qquad k=1,\dots,m_{\mathrm{hdr}}.
\]
On this grid, define the normalized cell masses
\begin{equation}
\hat p_{n,k}(\vb l)
=
\frac{\hat f_n(a^{\mathrm{hdr}}_k\mid \vb l)\,w_k}
{\sum_{r=1}^{m_{\mathrm{hdr}}}\hat f_n(a^{\mathrm{hdr}}_r\mid \vb l)\,w_r},
\qquad k=1,\dots,m_{\mathrm{hdr}}.
\label{eq:grid_weights}
\end{equation}

Equivalently, \(\{\hat p_{n,k}(\vb l)\}_{k=1}^{m_{\mathrm{hdr}}}\) defines a discrete distribution on the HDR grid
\(\{a^{\mathrm{hdr}}_1,\ldots,a^{\mathrm{hdr}}_{m_{\mathrm{hdr}}}\}\) (or, equivalently, on the partition
\(\{I_k\}_{k=1}^{m_{\mathrm{hdr}}}\)). Let \(\hat A_{\vb l}\) denote a discrete random variable such that
\[
P\!\left(\hat A_{\vb l}=a^{\mathrm{hdr}}_k \mid \vb L=\vb l\right)=\hat p_{n,k}(\vb l),
\qquad k=1,\ldots,m_{\mathrm{hdr}}.
\]
Define the corresponding estimated density-value random variable by
\[
\hat Z_{\vb l}:=\hat f_n\!\left(\hat A_{\vb l}\mid \vb l\right).
\]
Its conditional distribution function is
\[
P\!\left(\hat Z_{\vb l}\le t\mid \vb L=\vb l\right)
=
\sum_{k=1}^{m_{\mathrm{hdr}}}\hat p_{n,k}(\vb l)\,
\mathbbm 1\!\left\{\hat f_n(a^{\mathrm{hdr}}_k\mid \vb l)\le t\right\},
\qquad t\in\mathbb R.
\]

Then \(\hat f_{n,\alpha}(\vb l)\) is the \((1-\alpha)\)-quantile of \(\hat Z_{\vb l}\):
\begin{equation}
\hat f_{n,\alpha}(\vb l)
=
\inf\Big\{t\in\mathbb R:\ \sum_{k=1}^{m_{\mathrm{hdr}}}\hat p_{n,k}(\vb l)\,
\mathbbm 1\!\big\{\hat f_n(a^{\mathrm{hdr}}_k\mid \vb l)\le t\big\}\ge 1-\alpha\Big\}.
\label{eq:fa_quantile_hat}
\end{equation}
The estimated supported set on the HDR grid is
\begin{equation}
\hat{\mathcal A}^{\,\mathrm{hdr}}_{n,\alpha}(\vb l)
=
\Big\{a^{\mathrm{hdr}}_k\in\mathcal A:\ \hat f_n(a^{\mathrm{hdr}}_k\mid \vb l)\ge \hat f_{n,\alpha}(\vb l)\Big\}.
\label{eq:hdr_hat_closed_def}
\end{equation}

\begin{remark}[Quality of the grid approximation]
\label{rem:B_grid_quality}
The quality of the HDR approximation depends on both density-estimation error and
grid discretization error.

Let
\[
\Delta_n(\vb l):=\sup_{a\in\mathcal A}\big|\hat f_n(a\mid \vb l)-f_0(a\mid \vb l)\big|,
\]
and let \(\delta_m:=\max_{1\le k\le m_{\mathrm{hdr}}} w_k\) denote the maximum cell
width of the HDR-grid partition. Under compact support and regularity of
\(f_0(\cdot\mid \vb l)\) (e.g., continuity in \(a\)), the grid-based
approximation error in cutoff and membership is controlled by these two terms: a
stochastic part (\(\Delta_n\)) and a deterministic discretization part
(\(\delta_m\)). Heuristically,
\[
\text{HDR approximation error}
\;\lesssim\;
\Delta_n(\vb l)+\delta_m,
\]
uniformly in \(\vb l\) under corresponding uniform conditions.

Hence, approximation improves when (i) \(\hat f_n\) is accurate and (ii) the grid
is sufficiently fine. In practice, one can assess adequacy by a simple
grid-refinement check: recompute \(\hat f_{n,\alpha}\), \(\hat\tau_{n,\alpha}\),
and the feasible curve on progressively finer grids (e.g., halve \(\delta_m\)) and
verify numerical stability. Large changes under refinement indicate that the
current grid is too coarse in regions relevant to the estimated cutoff.
\end{remark}

\section{Proof of Convergence of the Feasible Curve}\label{apd:proof_fdrc}

\subsection{Setup, notation, and target}\label{apd:proof_fdrc:setup}
Let $O=(\vb L,A,Y)\sim P_0$ and let $\{O_i\}_{i=1}^n$ be i.i.d.\ draws from $P_0$. Denote by
$f_0(a\mid \vb l)$ the conditional density of $A$ given $\vb L=\vb l$ and by
$\mu_0(a,\vb l)=\E_0(Y\mid A=a,\vb L=\vb l)$ the outcome regression.

Fix a support level $\alpha\in(0,1)$. For each $\vb l$, let
\[
\mathcal A_\alpha(\vb l;P_0)=\{a\in\mathcal A:\ f_0(a\mid \vb l)\ge f_\alpha(\vb l)\}
\]
be the $\alpha$-HDR region defined in Appendix~\ref{app:fa_quantile}. Define the feasible intervention operator
\[
d_\alpha(a,\vb l)=
\begin{cases}
a, & a\in \mathcal A_\alpha(\vb l;P_0),\\
\Pi_{\mathcal A_\alpha(\vb l;P_0)}(a), & a\notin \mathcal A_\alpha(\vb l;P_0),
\end{cases}
\]
where $\Pi_T(\cdot)$ is the (single-valued) projection onto a nonempty closed set $T\subset\mathbb R$ using a
fixed deterministic measurable tie-breaking convention.

Under the identification conditions stated in the main text, the feasible curve can be written as
\begin{equation}\label{eq:apd_target_functional}
m^{\text{feasible}}(a)
=
\E_0\left[\mu_0\left(d_\alpha(a,\vb L),\vb L\right)\right],
\qquad a\in\mathcal A.
\end{equation}

\subsection{Estimator induced by Algorithm~\ref{alg:feasible}}\label{apd:proof_fdrc:estimator}
Algorithm~\ref{alg:feasible} yields:
\begin{itemize}
\item an estimator $\hat f_n(a\mid \vb l)$ of $f_0(a\mid \vb l)$;
\item for each $\vb l$, an estimated HDR cutoff $\hat f_{\alpha,n}(\vb l)$ (Appendix~\ref{app:fa_quantile});
\item an estimated supported set $\hat{\mathcal A}_{\alpha,n}(\vb l)$, defined either on $\mathcal A$ or on a grid,
      and in either case as a closed upper level set at the estimated cutoff;
\item an estimated feasible rule $\hat d_{\alpha,n}(a,\vb l)$ defined by replacing $\mathcal A_\alpha(\vb l;P_0)$
      with $\hat{\mathcal A}_{\alpha,n}(\vb l)$ and using the same tie-breaking convention;
\item an estimator $\hat\mu_n(a,\vb l)$ of $\mu_0(a,\vb l)$.
\end{itemize}
The plug-in estimator of the feasible curve is
\begin{equation}\label{eq:apd_plugin}
\hat m^{\text{feasible}}_n(a)
=
\frac{1}{n}\sum_{i=1}^n
\hat\mu_n\left(\hat d_{\alpha,n}(a,\vb L_i),\vb L_i\right),
\qquad a\in\mathcal A.
\end{equation}
(The proof below is pointwise in $a$; convergence on a finite grid $\{a_1,\dots,a_m\}$ follows immediately.)

\subsection{Assumptions}\label{apd:proof_fdrc:assumptions}
We state sufficient conditions for pointwise consistency of $\hat m^{\text{feasible}}_n(a)$.

\begin{enumerate}[label=\textbf{(A\arabic*)}]
\item \textbf{Identification assumptions.} The identification conditions for the feasible estimand in the main text
      hold (consistency, conditional exchangeability, and well-definedness of the projection rule).

\item \textbf{Compact treatment space.} $\mathcal A\subset\mathbb R$ is compact.

\item \textbf{Well-defined supported sets.} For $P_0$-almost every $\vb l$, the set
      $\mathcal A_\alpha(\vb l;P_0)=\{a\in\mathcal A: f_0(a\mid \vb l)\ge f_\alpha(\vb l)\}$ is nonempty and closed.

\item \textbf{No mass on the cutoff boundary.} For $P_0$-almost every $\vb l$,
\[
P_0\left(f_0(A\mid \vb l)=f_\alpha(\vb l)\,\middle|\,\vb L=\vb l\right)=0.
\]

\item \textbf{Uniform consistency of the density estimator.} For $P_0$-almost every $\vb l$,
\[
\sup_{a\in\mathcal A}\big|\hat f_n(a\mid \vb l)-f_0(a\mid \vb l)\big|\ \stackrel{P}{\to}\ 0.
\]

\item \textbf{Consistency of the estimated cutoff.} For $P_0$-almost every $\vb l$,
\[
\hat f_{\alpha,n}(\vb l)\ \stackrel{P}{\to}\ f_\alpha(\vb l).
\]

\item \textbf{Level-set regularity.} For $P_0$-almost every $\vb l$, the map $a\mapsto f_0(a\mid \vb l)$ is
      continuous on $\mathcal A$.

\item \textbf{Outcome regression consistency and boundedness.}
For \(P_0\)-almost every \(\vb l\),
\[
\sup_{a\in\mathcal A}\big|\hat\mu_n(a,\vb l)-\mu_0(a,\vb l)\big|\ \stackrel{P}{\to}\ 0.
\]
Moreover, \(Y\) is bounded almost surely (e.g., \(Y\in[0,1]\)), so that
\(\sup_{a\in\mathcal A}|\mu_0(a,\vb L)|\le \|Y\|_\infty<\infty\) and
\(\sup_{a\in\mathcal A}|\hat\mu_n(a,\vb L)|\le \|Y\|_\infty<\infty\) with probability tending to one.

\end{enumerate}

\begin{remark}[Sufficient conditions for Assumption~\textbf{(A6)}]\label{rem:A6_holds}
Assumption~\textbf{(A6)} requires \(\hat f_{\alpha,n}(\vb l)\stackrel{P}{\to} f_\alpha(\vb l)\), where
\(f_\alpha(\vb l)\) is the \((1-\alpha)\)-quantile of \(Z_{\vb l}=f_0(A\mid \vb l)\) defined in
\eqref{eq:fa_quantile_pop}. This is a standard quantile-consistency requirement. A convenient set of sufficient
conditions is as follows: for \(P_0\)-almost every \(\vb l\),
\begin{enumerate}[label=(\roman*)]
\item \emph{Uniform density consistency:}
\(\sup_{a\in\mathcal A}\lvert \hat f_n(a\mid \vb l)-f_0(a\mid \vb l)\rvert \stackrel{P}{\to}0\)
(as in Assumption~\textbf{(A5)}).
\item \emph{No atom at the cutoff:} the conditional distribution function of \(Z_{\vb l}\) is continuous at
\(f_\alpha(\vb l)\), which is implied by Assumption~\textbf{(A4)}.
\item \emph{Cutoff computed as a quantile of an estimated density-value distribution:}
\(\hat f_{\alpha,n}(\vb l)\) is obtained as the \((1-\alpha)\)-quantile of an estimated analogue
\(\hat Z_{\vb l}\) of \(Z_{\vb l}\) that is asymptotically equivalent to the population construction in
Appendix~\ref{app:fa_quantile}. For example, one may take \(\hat Z_{\vb l}=\hat f_n(A\mid \vb l)\) with
\(A\sim \hat f_n(\cdot\mid \vb l)\), approximated by Monte Carlo draws; or one may use the grid-based analogue in
Appendix~\ref{app:fa_quantile:sample}, where \(\hat Z_{\vb l}=\hat f_n(\hat A_{\vb l}\mid \vb l)\) and
\(P(\hat A_{\vb l}=a^{\mathrm{hdr}}_k\mid \vb L=\vb l)=\hat p_{n,k}(\vb l)\).
\end{enumerate}
Under (i)--(iii), the conditional distribution function of \(\hat Z_{\vb l}\) converges to that of \(Z_{\vb l}\)
at the cutoff, and continuity at \(f_\alpha(\vb l)\) yields
\(\hat f_{\alpha,n}(\vb l)\stackrel{P}{\to} f_\alpha(\vb l)\).

For the grid-based construction, an additional discretization condition is needed to ensure that the induced
distribution of \(\hat A_{\vb l}\) approximates \(A\mid \vb L=\vb l\). A sufficient condition is that the maximum
cell width \(\delta_m:=\max_{1\le k\le m_{\mathrm{hdr}}} w_k\) is small (and, in an asymptotic analysis,
\(\delta_m\to 0\)) so that the Riemann-sum approximation error in
\(\sum_{k=1}^{m_{\mathrm{hdr}}}\hat f_n(a^{\mathrm{hdr}}_k\mid \vb l)w_k\) and the resulting weighted empirical
distribution of \(\hat Z_{\vb l}\) is negligible.
\end{remark}

\subsection{Intermediate lemmas}\label{apd:proof_fdrc:lemmas}

\noindent\textbf{Notation (for Lemmas~\ref{apd:proof_fdrc:lemma1}--\ref{apd:proof_fdrc:lemma_rule}).}
Fix $\alpha\in(0,1)$ and a covariate value $\vb l$. Suppress $(\alpha,\vb l)$ and write
\[
S:=\mathcal A_\alpha(\vb l;P_0),
\qquad
S_n:=\hat{\mathcal A}_{\alpha,n}(\vb l).
\]
Both $S$ and $S_n$ are nonempty closed subsets of $\mathcal A\subset\mathbb R$.

For $x\in\mathbb R$ and a nonempty set $T\subset\mathbb R$, define
\[
\dist(x,T):=\inf_{t\in T}|x-t|.
\]
For two nonempty compact sets $T_1,T_2\subset\mathbb R$, the Hausdorff distance is
\[
d_H(T_1,T_2):=\max\Big\{\sup_{t\in T_1}\dist(t,T_2),\ \sup_{t\in T_2}\dist(t,T_1)\Big\}.
\]
For a nonempty closed set $T\subset\mathbb R$, define $\Pi_T(x)\in \arg\min_{t\in T}|x-t|$ using a fixed,
deterministic measurable tie-breaking convention.

Then
\[
d_\alpha(a,\vb l)=a\,\mathbbm 1\{a\in S\}+\Pi_S(a)\,\mathbbm 1\{a\notin S\},
\qquad
\hat d_{\alpha,n}(a,\vb l)=a\,\mathbbm 1\{a\in S_n\}+\Pi_{S_n}(a)\,\mathbbm 1\{a\notin S_n\}.
\]

\medskip

\bmsubsubsection{Lemma 1 (Stability of supported-set membership)}\label{apd:proof_fdrc:lemma1}
Fix $\alpha\in(0,1)$. Under \textbf{(A5)}--\textbf{(A6)}, for $P_0$-almost every $\vb l$ and any fixed
$a\in\mathcal A$ with $f_0(a\mid \vb l)\neq f_\alpha(\vb l)$,
\[
\mathbbm 1\left\{\hat f_n(a\mid \vb l) \ge \hat f_{\alpha,n}(\vb l)\right\}
\ \stackrel{P}{\to}\
\mathbbm 1\left\{f_0(a\mid \vb l) \ge f_\alpha(\vb l)\right\}.
\]
Equivalently (away from boundary points),
\[
\mathbbm 1\left\{a\in S_n(\vb l)\right\}
\ \stackrel{P}{\to}\
\mathbbm 1\left\{a\in S(\vb l)\right\}.
\]

\textit{Proof.}
Fix \(\vb l\) and a point \(a\in\mathcal A\) such that
\[
\Delta(\vb l):=\big|f_0(a\mid \vb l)-f_\alpha(\vb l)\big|>0.
\]
Set \(\varepsilon(\vb l):=\Delta(\vb l)/3\), and define the event
\[
E_n(\vb l):=
\left\{
\sup_{u\in\mathcal A}\big|\hat f_n(u\mid \vb l)-f_0(u\mid \vb l)\big|<\varepsilon(\vb l)
\right\}
\cap
\left\{
\big|\hat f_{\alpha,n}(\vb l)-f_\alpha(\vb l)\big|<\varepsilon(\vb l)
\right\}.
\]
By Assumptions \textbf{(A5)}--\textbf{(A6)}, \(P_0\big(E_n(\vb l)\big)\to 1\).

On \(E_n(\vb l)\), we have
\[
\begin{aligned}
&\Big|
\big(\hat f_n(a\mid \vb l)-\hat f_{\alpha,n}(\vb l)\big)
-
\big(f_0(a\mid \vb l)-f_\alpha(\vb l)\big)
\Big|\\
&\qquad\le
\big|\hat f_n(a\mid \vb l)-f_0(a\mid \vb l)\big|
+
\big|\hat f_{\alpha,n}(\vb l)-f_\alpha(\vb l)\big|
<
2\varepsilon(\vb l)
=
\frac{2}{3}\Delta(\vb l).
\end{aligned}
\]
Hence the sign of
\(\hat f_n(a\mid \vb l)-\hat f_{\alpha,n}(\vb l)\)
must match the sign of
\(f_0(a\mid \vb l)-f_\alpha(\vb l)\):
\begin{itemize}
\item If \(f_0(a\mid \vb l)>f_\alpha(\vb l)\), then
\[
\hat f_n(a\mid \vb l)-\hat f_{\alpha,n}(\vb l)
>
\Delta(\vb l)-\frac{2}{3}\Delta(\vb l)
=
\frac{1}{3}\Delta(\vb l)>0,
\]
so \(\hat f_n(a\mid \vb l)\ge \hat f_{\alpha,n}(\vb l)\).

\item If \(f_0(a\mid \vb l)<f_\alpha(\vb l)\), then
\[
\hat f_n(a\mid \vb l)-\hat f_{\alpha,n}(\vb l)
<
-\Delta(\vb l)+\frac{2}{3}\Delta(\vb l)
=
-\frac{1}{3}\Delta(\vb l)<0,
\]
so \(\hat f_n(a\mid \vb l)<\hat f_{\alpha,n}(\vb l)\).
\end{itemize}
Therefore, on \(E_n(\vb l)\),
\[
\mathbbm 1\left\{\hat f_n(a\mid \vb l)\ge \hat f_{\alpha,n}(\vb l)\right\}
=
\mathbbm 1\left\{f_0(a\mid \vb l)\ge f_\alpha(\vb l)\right\}.
\]
So indicator disagreement implies \(E_n(\vb l)^c\), and thus
\[
\begin{aligned}
&P_0\left(
\mathbbm 1\{\hat f_n(a\mid \vb l)\ge \hat f_{\alpha,n}(\vb l)\}
\neq
\mathbbm 1\{f_0(a\mid \vb l)\ge f_\alpha(\vb l)\}
\right)\\
&\qquad\le P_0\big(E_n(\vb l)^c\big)\to 0.
\end{aligned}
\]
This proves
\[
\mathbbm 1\left\{\hat f_n(a\mid \vb l)\ge \hat f_{\alpha,n}(\vb l)\right\}
\stackrel{P}{\to}
\mathbbm 1\left\{f_0(a\mid \vb l)\ge f_\alpha(\vb l)\right\}.
\]
\hfill\(\square\)

\medskip

\bmsubsubsection{Lemma 2 (Pointwise convergence of the feasible rule)}\label{apd:proof_fdrc:lemma_rule}
Assume \textbf{(A2)}--\textbf{(A7)} and use a deterministic measurable tie-breaking convention to make
$\Pi_T(\cdot)$ single-valued for every nonempty closed $T\subset\mathbb R$, fixed across $n$. Then for each fixed
$a\in\mathcal A$, for $P_0$-almost every $\vb l$,
\[
\hat d_{\alpha,n}(a,\vb l)\ \stackrel{P}{\to}\ d_\alpha(a,\vb l).
\]

\textit{Proof.}
Fix \(a\in\mathcal A\) and \(\vb l\) in a probability-one set where Assumptions \textbf{(A2)}--\textbf{(A7)} hold.

By Lemma~\ref{apd:proof_fdrc:lemma1}, away from the boundary case \(f_0(a\mid \vb l)=f_\alpha(\vb l)\),
\[
\mathbbm 1\{a\in S_n\}\stackrel{P}{\to}\mathbbm 1\{a\in S\}.
\]
Hence it remains to prove projection convergence on the complement event \(a\notin S\):
\begin{equation}\label{eq:proj_conv_goal_rewrite}
\Pi_{S_n}(a)\ \stackrel{P}{\to}\ \Pi_S(a).
\end{equation}

We first show that \(S_n\) converges to \(S\) in Hausdorff distance.

Let
\[
\partial S:=\{u\in\mathcal A:\ f_0(u\mid \vb l)=f_\alpha(\vb l)\}.
\]
For \(\eta>0\), define
\[
U_\eta:=\{u\in\mathcal A:\dist(u,\partial S)\le \eta\},
\qquad
K_\eta:=\mathcal A\setminus U_\eta.
\]
By \textbf{(A2)}, \(\mathcal A\) is compact, so \(K_\eta\) is compact. By continuity of \(u\mapsto f_0(u\mid \vb l)\)
(Assumption~\textbf{(A7)}), the continuous function
\(u\mapsto |f_0(u\mid \vb l)-f_\alpha(\vb l)|\) attains its minimum on \(K_\eta\). Since \(K_\eta\cap\partial S=\varnothing\),
this minimum is strictly positive:
\[
\delta_\eta(\vb l):=\min_{u\in K_\eta}|f_0(u\mid \vb l)-f_\alpha(\vb l)|>0.
\]

Define
\[
E_{n,\eta}(\vb l):=
\Big\{\sup_{u\in\mathcal A}|\hat f_n(u\mid \vb l)-f_0(u\mid \vb l)|<\delta_\eta(\vb l)/3\Big\}
\cap
\Big\{|\hat f_{\alpha,n}(\vb l)-f_\alpha(\vb l)|<\delta_\eta(\vb l)/3\Big\}.
\]
By \textbf{(A5)}--\textbf{(A6)}, \(P_0(E_{n,\eta}(\vb l))\to1\).

Now fix any \(u\in K_\eta\). On \(E_{n,\eta}(\vb l)\),
\[
\Big|
\big(\hat f_n(u\mid \vb l)-\hat f_{\alpha,n}(\vb l)\big)
-
\big(f_0(u\mid \vb l)-f_\alpha(\vb l)\big)
\Big|
<\frac{2}{3}\delta_\eta(\vb l).
\]
Because \(|f_0(u\mid \vb l)-f_\alpha(\vb l)|\ge \delta_\eta(\vb l)\), this perturbation cannot flip the sign; therefore
\[
\mathbbm 1\{\hat f_n(u\mid \vb l)\ge \hat f_{\alpha,n}(\vb l)\}
=
\mathbbm 1\{f_0(u\mid \vb l)\ge f_\alpha(\vb l)\},
\qquad \forall u\in K_\eta.
\]
Equivalently,
\[
S_n\cap K_\eta=S\cap K_\eta
\qquad\text{on }E_{n,\eta}(\vb l).
\]
So any discrepancy between \(S_n\) and \(S\) can occur only inside \(U_\eta\), i.e., within distance \(\eta\) of \(\partial S\).
Hence
\[
d_H(S_n,S)\le \eta
\qquad\text{on }E_{n,\eta}(\vb l).
\]
Since \(P_0(E_{n,\eta}(\vb l))\to1\), for each fixed \(\eta>0\),
\[
P_0\big(d_H(S_n,S)>\eta\big)\to0,
\]
which implies \(d_H(S_n,S)\stackrel{P}{\to}0\).

Now assume \(a\notin S\) and define
\[
p_n:=\Pi_{S_n}(a),\qquad p:=\Pi_S(a).
\]
Since \(S_n,S\) are nonempty closed subsets of compact \(\mathcal A\subset\mathbb R\), they are compact. Hence:
(i) the minima \(\min_{t\in S_n}|a-t|\) and \(\min_{t\in S}|a-t|\) are attained, so \(p_n,p\) are well-defined;
(ii) all \(p_n\in\mathcal A\), so \(\{p_n\}\) is relatively compact.

For fixed \(a\),
\[
\big|\dist(a,S_n)-\dist(a,S)\big|
\le d_H(S_n,S)\stackrel{P}{\to}0.
\]
Therefore
\[
|a-p_n|=\dist(a,S_n)\stackrel{P}{\to}\dist(a,S)=|a-p|.
\]

It remains to pass from distance convergence to projector convergence.  
Consider any subsequence \(n_k\). By compactness of \(\mathcal A\), \(\{p_{n_k}\}\subset\mathcal A\) has a further subsequence
\(p_{n_{k_r}}\to q\in\mathcal A\). Because \(p_{n_{k_r}}\in S_{n_{k_r}}\) and \(d_H(S_{n_{k_r}},S)\to 0\), every such limit point must lie in \(S\), so \(q\in S\). Also,
\[
|a-q|
=
\lim_{r\to\infty}|a-p_{n_{k_r}}|
=
\lim_{r\to\infty}\dist(a,S_{n_{k_r}})
=
\dist(a,S).
\]
Hence \(q\in\arg\min_{t\in S}|a-t|\).

By the fixed deterministic tie-breaking convention, \(\Pi_S(a)\) is a unique selected minimizer. The same convention is used for every \(S_n\). Therefore any subsequential limit \(q\) of \(\{p_n\}\) must equal this selected minimizer \(p=\Pi_S(a)\). Since every subsequence has only this possible limit, we conclude
\[
p_n=\Pi_{S_n}(a)\stackrel{P}{\to}\Pi_S(a)=p.
\]
This proves \eqref{eq:proj_conv_goal_rewrite}.

\medskip

\subsection{Identification of the Feasible Dose--Response Curve}
\label{apd:proof_fdrc_identification}

For each fixed \(a\in\mathcal A\), define
\[
d_a(\vb l):=d_{0,\alpha}(a,\vb l).
\]
By Assumption 3 in Estimand 3, \(d_a(\vb l)\) is well defined for \(P_{0,\vb L}\)-almost every \(\vb l\), and \(\vb l\mapsto d_a(\vb l)\) is measurable.

Starting from the definition,
\[
m_0^{\mathrm{feasible}}(a)
=
\E_0\!\left\{Y^{d_a(\cdot)}\right\}
=
\E_0\!\left[\E_0\!\left\{Y^{d_a(\cdot)}\mid \vb L\right\}\right].
\]
Fix \(\vb l\). Since \(d_a(\vb l)\) is a fixed treatment level, conditional exchangeability implies
\[
\E_0\!\left\{Y^{d_a(\vb l)}\mid \vb L=\vb l\right\}
=
\E_0\!\left\{Y^{d_a(\vb l)}\mid A=d_a(\vb l),\,\vb L=\vb l\right\}.
\]
By consistency, on \(\{A=d_a(\vb l)\}\), \(Y^{d_a(\vb l)}=Y\), hence
\[
\E_0\!\left\{Y^{d_a(\vb l)}\mid A=d_a(\vb l),\,\vb L=\vb l\right\}
=
\E_0\!\left\{Y\mid A=d_a(\vb l),\,\vb L=\vb l\right\}.
\]
Therefore, for \(P_{0,\vb L}\)-almost every \(\vb l\),
\[
\E_0\!\left\{Y^{d_a(\cdot)}\mid \vb L=\vb l\right\}
=
\E_0\!\left\{Y\mid A=d_a(\vb l),\,\vb L=\vb l\right\}.
\]
Integrating over \(\vb L\) yields
\[
m_0^{\mathrm{feasible}}(a)
=
\E_0\!\left[
\E_0\!\left\{Y\mid A=d_{0,\alpha}(a,\vb L),\,\vb L\right\}
\right].
\]
This establishes identification.

\paragraph{Interpretation relative to positivity.}
Unlike a static intervention that fixes treatment at \(a\) for all individuals, the feasible rule assigns \(d_{0,\alpha}(a,\vb L)\), which is constructed to lie in \(\mathcal A_\alpha(\vb L;P_0)\) almost surely. Thus, identification is anchored to support at the assigned feasible level, rather than requiring uniform support at the original target \(a\) in every covariate stratum.

\section{Main theorem and proof}\label{apd:proof_fdrc:main}

\subsection{Theorem (Pointwise convergence of the feasible plug-in estimator)}\label{apd:proof_fdrc:thm}
Under Assumptions \textbf{(A1)}--\textbf{(A7)} and \textbf{(A8)}(A) (uniform outcome-regression consistency and bounded \(Y\)),
for each fixed \(a\in\mathcal A\),
\[
\hat m^{\text{feasible}}_n(a)\ \stackrel{P}{\to}\ m^{\text{feasible}}(a).
\]

\textit{Proof.}
Fix \(a\in\mathcal A\). Recall
\[
m^{\text{feasible}}(a)=\E_0\left[\mu_0\left(d_\alpha(a,\vb L),\vb L\right)\right],
\qquad
\hat m^{\text{feasible}}_n(a)=\frac1n\sum_{i=1}^n \hat\mu_n\left(\hat d_{\alpha,n}(a,\vb L_i),\vb L_i\right).
\]
Define
\[
\psi_0(\vb l):=\mu_0(d_\alpha(a,\vb l),\vb l),
\qquad
\hat\psi_n(\vb l):=\hat\mu_n(\hat d_{\alpha,n}(a,\vb l),\vb l).
\]
Then the estimation error admits the two-term decomposition
\[
\hat m^{\text{feasible}}_n(a)-m^{\text{feasible}}(a)
=
\underbrace{\Big\{\mathbb P_n\hat\psi_n-P_0\hat\psi_n\Big\}}_{(I)}
+
\underbrace{\Big\{P_0\hat\psi_n-P_0\psi_0\Big\}}_{(II)},
\]
where \(\mathbb P_n g:=n^{-1}\sum_{i=1}^n g(\vb L_i)\) and \(P_0 g:=\E_0\{g(\vb L)\}\).

\medskip
\noindent\emph{Step 1: \((II)\stackrel{P}{\to}0\).}
For any \(\vb l\),
\begin{align*}
\big|\hat\psi_n(\vb l)-\psi_0(\vb l)\big|
&=
\big|\hat\mu_n(\hat d_{\alpha,n}(a,\vb l),\vb l)-\mu_0(d_\alpha(a,\vb l),\vb l)\big|\\
&\le
\underbrace{\big|\hat\mu_n(\hat d_{\alpha,n}(a,\vb l),\vb l)-\mu_0(\hat d_{\alpha,n}(a,\vb l),\vb l)\big|}_{(A)}
+
\underbrace{\big|\mu_0(\hat d_{\alpha,n}(a,\vb l),\vb l)-\mu_0(d_\alpha(a,\vb l),\vb l)\big|}_{(B)}.
\end{align*}
Term (A) is bounded by the uniform regression error:
\[
(A)\le \sup_{u\in\mathcal A}\big|\hat\mu_n(u,\vb l)-\mu_0(u,\vb l)\big|
\ \stackrel{P}{\to}\ 0
\quad\text{for }P_0\text{-a.e.\ }\vb l,
\]
by Assumption \textbf{(A8)}(A).
For (B), Lemma~\ref{apd:proof_fdrc:lemma_rule} yields
\(\hat d_{\alpha,n}(a,\vb l)\stackrel{P}{\to} d_\alpha(a,\vb l)\) for \(P_0\)-a.e.\ \(\vb l\). If, in addition,
\(u\mapsto \mu_0(u,\vb l)\) is continuous on the compact \(\mathcal A\) for \(P_0\)-a.e.\ \(\vb l\), then the
continuous mapping theorem gives \((B)\stackrel{P}{\to}0\) for \(P_0\)-a.e.\ \(\vb l\).
Hence \(\hat\psi_n(\vb l)\to \psi_0(\vb l)\) in probability for \(P_0\)-a.e.\ \(\vb l\).

Since \(Y\) is bounded (Assumption \textbf{(A8)}(A)), both \(\mu_0(\cdot,\vb L)\) and \(\hat\mu_n(\cdot,\vb L)\)
take values in a bounded range, and thus \(|\hat\psi_n(\vb L)-\psi_0(\vb L)|\le 2\|Y\|_\infty\) almost surely.
Dominated convergence implies
\[
P_0\big|\hat\psi_n-\psi_0\big|\ \to\ 0,
\]
so \((II)=P_0(\hat\psi_n-\psi_0)\stackrel{P}{\to}0\).

\medskip
\noindent\emph{Step 2: \((I)\stackrel{P}{\to}0\).}
At this step, we require a condition ensuring that the empirical average of the (random) function \(\hat\psi_n\)
tracks its population mean \(P_0\hat\psi_n\). This can be obtained either by sample splitting/cross-fitting or by
a weak additional assumption stated below.

\smallskip
\noindent\emph{Case 2a (sample splitting / cross-fitting).}
Suppose \(\hat\psi_n\) is fit using a training sample that is independent of the evaluation sample
\(\{\vb L_i\}_{i=1}^n\) used in \(\mathbb P_n\hat\psi_n\) (e.g., by sample splitting, or by cross-fitting with
foldwise evaluation). Conditional on the training data, \(\hat\psi_n(\cdot)\) is fixed, while the evaluation-fold
covariates remain i.i.d.\ from \(P_0\). Therefore, conditional on the training data,
\[
\E_0\!\left[\mathbb P_n\hat\psi_n-P_0\hat\psi_n\ \middle|\ \hat\psi_n\right]=0
\quad\text{and}\quad
\Var_0\!\left(\mathbb P_n\hat\psi_n-P_0\hat\psi_n\ \middle|\ \hat\psi_n\right)
=
\frac{1}{n}\Var_0\!\left(\hat\psi_n(\vb L)\ \middle|\ \hat\psi_n\right).
\]
By boundedness, \(|\hat\psi_n(\vb L)|\le \|Y\|_\infty\), hence
\(\Var_0(\hat\psi_n(\vb L)\mid \hat\psi_n)\le \|Y\|_\infty^2\). Chebyshev's inequality (conditionally on
\(\hat\psi_n\)) gives \((I)\stackrel{P}{\to}0\).

\smallskip
\noindent\emph{Case 2b (no splitting; weak additional assumption).}
If \(\hat\psi_n\) is fitted on the same sample used in \(\mathbb P_n\), we impose the following weak condition:
\begin{equation}\label{eq:weak_empirical_process}
(\mathbb P_n-P_0)\hat\psi_n\ \stackrel{P}{\to}\ 0.
\end{equation}

Under \eqref{eq:weak_empirical_process}, we
immediately obtain \((I)\stackrel{P}{\to}0\).

\medskip
Combining Step 1 with either Case 2a or Case 2b yields
\(\hat m^{\text{feasible}}_n(a)-m^{\text{feasible}}(a)\stackrel{P}{\to}0\), completing the proof.
\hfill\(\square\)

\section{Additional Estimation Notes: Inference and First-Stage Estimation}
\label{apd:estimation_additional_notes}

\begin{enumerate}[label=\arabic*)]
\setcounter{enumi}{3}

\item
Valid statistical inference via the nonparametric bootstrap (Algorithm~\ref{alg:feasible}, Step~7) is most transparent for estimators that are regular and asymptotically linear (RAL) for a \emph{fixed} target parameter, under standard regularity conditions \cite{bickel1981some,van2000asymptotic,van1996weak,kosorok2008introduction}. These conditions hold, for example, under correctly specified low-dimensional parametric outcome models, and can also hold for certain flexible estimators such as the Highly Adaptive Lasso (HAL) \cite{benkeser2016highly} in pathwise differentiable settings \cite{ButzinDozier2024,van2023higher}. The standard CDRC, viewed as an infinite-dimensional curve-valued parameter, is generally not pathwise differentiable, so uniform bootstrap validity for the \emph{entire} curve is not guaranteed in general. Nevertheless, pointwise inference at a fixed intervention level can be justified under stronger conditions that yield asymptotic linearity for that pointwise target (e.g., correct parametric specification and sufficient local overlap). Moreover, HAL plug-in estimators can yield pointwise asymptotic normality under undersmoothing, enabling delta-method inference based on the influence curve of the HAL coefficients \cite{shi2024hal}; related approaches have been used for g-computation variance estimation \cite{zivich2024empirical}.

\item
For the feasible and trimmed targets, the population estimands in \eqref{eqn:estimand_feasible} and \eqref{eqn:estimand_trim} depend on \(P_0\) through the conditional supported region \(\mathcal A_\alpha(\vb l;P_0)\), the induced feasible mapping \(d_{0,\alpha}(a,\vb l)\), and the non-overlap ratio \(\tau_{0,\alpha}(a)\) (cf.\ \eqref{eqn:nonoverlap}). In practice these objects are unknown and are replaced by plug-in estimates \(\hat{\mathcal A}^{\,\mathrm{hdr}}_{n,\alpha}(\vb l)\), \(\hat d_{n,\alpha}(a,\vb l)\), and \(\hat\tau_{n,\alpha}(a)\) (Algorithm~\ref{alg:feasible}, Steps~3--4 and 6b--6c). Consequently, \(\hat m^{\mathrm{feasible}}_{n,\alpha}\) and \(\hat m^{\mathrm{trim}}_{n,\alpha}\) incorporate an additional first-stage estimation step beyond \(\hat\mu_n\): the same sample is used both (i) to construct the supported set and the feasible/trimmed weights and (ii) to evaluate the resulting g-formula averages.

This reuse induces ``own-observation'' dependence: each observation influences the fitted density and supported set (and hence \(\hat d_{n,\alpha}\) and \(\hat t_{n,\alpha}\)) and then re-enters the final plug-in average. Such dependence does not preclude consistency when the empirical evaluation step behaves regularly, for example when \((\mathbb P_n-P_0)\hat\psi_n=o_P(1)\) for the fitted function \(\hat\psi_n(\vb l)=\hat\mu_n(\hat d_{n,\alpha}(a,\vb l),\vb l)\), or when sample splitting/cross-fitting is used. However, it complicates asymptotic linearity arguments and can affect bootstrap performance, particularly near the feasibility boundary where small perturbations of \(\hat f_n\) may change set membership or projection decisions. More generally, because the supported set enters through a non-smooth level-set operation and projection, bootstrap validity is less automatic than in smooth plug-in problems and may be sensitive to weak-overlap regions.

A standard remedy is sample splitting or cross-fitting: estimate \(\hat{\mathcal A}^{\,\mathrm{hdr}}_{n,\alpha}\) (and hence \(\hat d_{n,\alpha}\), \(\hat\tau_{n,\alpha}\), and \(\hat t_{n,\alpha}\)) on a training subsample (or training folds), and evaluate the corresponding g-formula on an independent validation subsample (or held-out folds). This separates, at least approximately, construction of the supported set/feasible rule from outcome-model evaluation, weakening own-observation dependence and simplifying large-sample arguments for inference. The trade-offs are additional implementation complexity and fold-specific supported sets and feasible rules; the resulting estimator averages fold-specific evaluations rather than relying on a single rule estimated from the full sample.

\item
In this work, we primarily use the full sample to estimate \(\hat{\mathcal A}^{\,\mathrm{hdr}}_{n,\alpha}\) (and hence \(\hat d_{n,\alpha}\), \(\hat\tau_{n,\alpha}\), and \(\hat t_{n,\alpha}\)) as well as the outcome regression \(\hat\mu_n\), yielding a single estimated supported set and a single estimated feasible rule for interpretation. Accordingly, the reported bootstrap intervals for \(\hat m_n^{\mathrm{standard}}\), \(\hat m_{n,\alpha}^{\mathrm{feasible}}\), and \(\hat m_{n,\alpha}^{\mathrm{trim}}\) should be interpreted as \emph{empirical} measures of uncertainty for the full estimation pipeline, rather than as intervals with a general theoretical coverage guarantee across all settings and first-stage learners. We therefore interpret them most cautiously in regions where feasibility decisions are frequent (high \(\hat\tau_{n,\alpha}(a)\)) or where the feasible mapping shows substantial remapping, since these are precisely the regions where the estimator is most sensitive to first-stage perturbations.

Our simulation study (Appendix~\ref{apd:bst}) empirically evaluates interval performance with and without sample splitting. In the settings considered, bootstrap intervals are close to nominal levels in well-supported regions, whereas departures from nominal behavior are more apparent in poorly supported regions, consistent with increased sensitivity of estimation and inference under weak practical positivity.
\end{enumerate}

\section{Simulation 1: Bootstrap CI coverage probabilities}\label{apd:bst}

\begin{figure}[htbp]
  \centering
  \subfloat[No sample splitting.\label{fig:sim1_bs_nosplit}]{
    \includegraphics[width=1\textwidth]{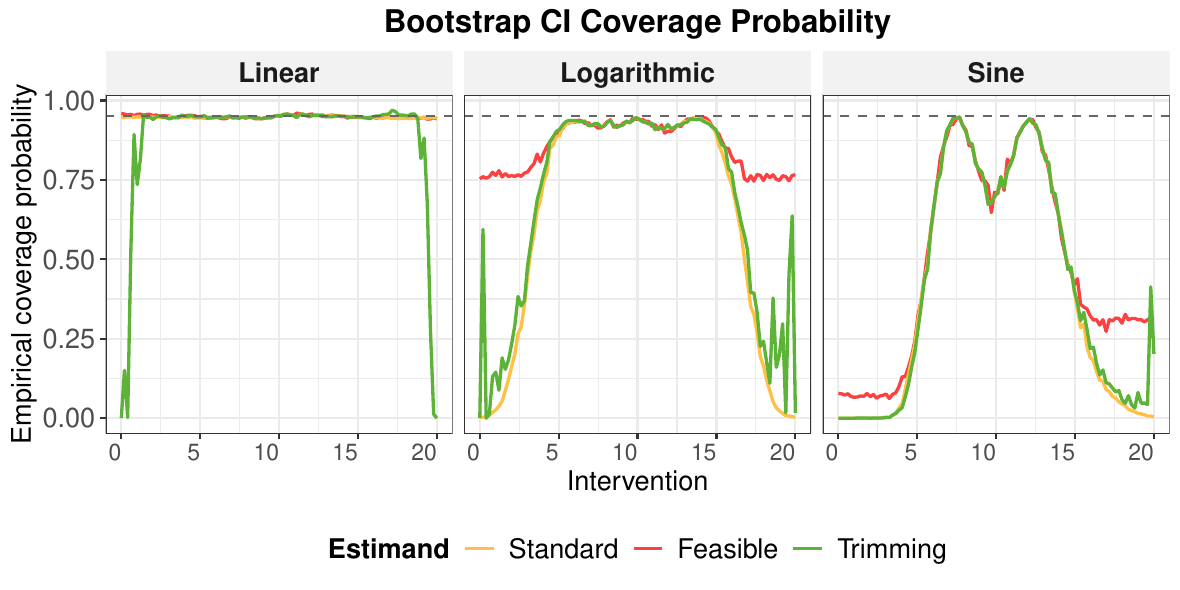}
  }\\[0.4em]
  \subfloat[With sample splitting.\label{fig:sim1_bs_split}]{
    \includegraphics[width=1\textwidth]{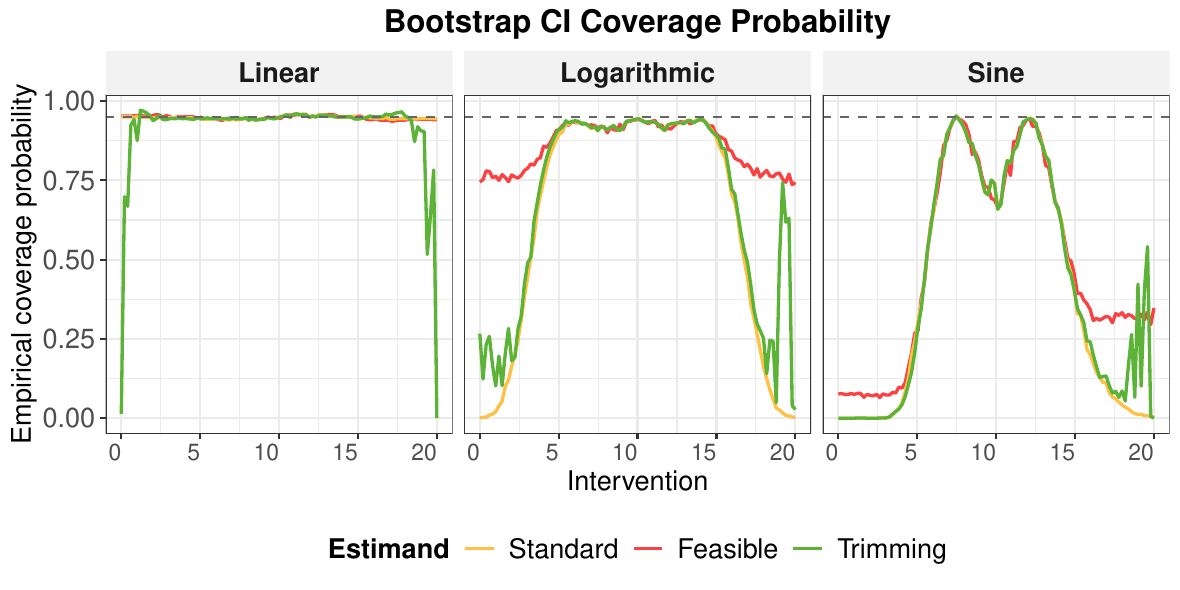}
  }
  \caption{
    \textbf{Bootstrap CI coverage in Simulation 1: comparison of sample splitting.}
    Empirical coverage of nominal 95\% nonparametric bootstrap confidence intervals for three pointwise curve estimators: the standard CDRC plug-in estimator \(\hat m_n^{\mathrm{standard}}\) (orange), the feasible curve estimator \(\hat m^{\mathrm{feasible}}_{n,\alpha}\) (red), and the trimmed curve estimator \(\hat m^{\mathrm{trim}}_{n,\alpha}\) (green), shown as a function of the target intervention value \(A\). Coverage is evaluated against the corresponding true target curves under the data-generating mechanism. Within each subfigure, panels correspond to different true outcome models (Linear, Logarithmic, Sine), and the horizontal dashed line indicates the nominal 95\% level. Panel (a) uses a full-sample bootstrap, where the first-stage quantities (supported set, feasible mapping, and trimming weights) and the g-formula evaluation are computed on the same resampled data. Panel (b) applies sample splitting within each bootstrap replicate, estimating the first-stage quantities on a training subset and evaluating the g-formula on held-out data. Differences between (a) and (b) reflect the potential impact of own-observation dependence from first-stage estimation, most notably near intervention values with limited practical overlap. In this simulation, however, coverage patterns are qualitatively similar with and without sample splitting.
    }
  \label{res:sim1_bs}
\end{figure}

See Figure~\ref{res:sim1_bs}.

\end{document}